\documentclass[a4paper,fleqn,usenatbib]{mnras}

\usepackage[T1]{fontenc}
\usepackage{ae,aecompl}
\usepackage{graphicx,longtable}
\usepackage{amsmath}
\usepackage{amssymb}
\usepackage{subfig}
\usepackage{soul}

\def\hequad{HE\,0435$-$1223}

\usepackage{natbib}
\usepackage[usenames]{color}

\title[External convergence in the field of \hequad]{H0LiCOW VIII. A weak lensing measurement of the external convergence in the field of the lensed quasar \hequad}

\author[O.~Tihhonova et al.]{\parbox{\textwidth}{
O. Tihhonova,$^{1}$\thanks{E-mail: olga.tihhonova@epfl.ch}
F. Courbin,$^{1}$
D. Harvey,$^{1}$
S.~Hilbert,$^{2,3}$
C.~E.~Rusu,$^{4,5}$
C.~D.~Fassnacht,$^{4}$
V.~Bonvin,$^{1}$
P.~J.~Marshall,$^{6}$
G.~Meylan,$^{1}$
D.~Sluse,$^{7}$
S.~H.~Suyu,$^{8,9,10}$
T.~Treu,$^{11}$
K.~C.~Wong$^{12}$
}
\\
\\
\parbox{\textwidth}{
	$^{1}$Laboratoire d'Astrophysique, Ecole Polytechnique
	F{\'e}d{\'e}rale de Lausanne (EPFL), Observatoire de Sauverny, CH-1290
	Versoix, Switzerland\\
	$^{2}$Exzellenzcluster Universe, Boltzmannstr. 2, 85748 Garching, Germany \\
	$^{3}$Ludwig-Maximilians-Universit{\"a}t, Universit{\"a}ts-Sternwarte, Scheinerstr. 1, 81679 M{\"u}nchen, Germany \\
	$^{4}$Department of Physics, University of California, Davis, CA
	95616, USA\\
	$^{5}$Subaru Telescope, 650 N Aohoku Pl, Hilo, HI 96720, USA\\
	$^{6}$Kavli Institute for Particle Astrophysics and Cosmology,
	Stanford University, 452 Lomita Mall, Stanford, CA 94035, USA\\
	$^{7}$STAR Institute, Quartier Agora - All\'ee du six Ao\^ut, 19c
	B-4000 Li\`ege, Belgium\\
	$^{8}$Max-Planck-Institut f{\"u}r Astrophysik, Karl-Schwarzschild-Str.~1, 85748 Garching, Germany\\
	$^{9}$Institute of Astronomy and Astrophysics, Academia Sinica, P.O.~Box 23-141, Taipei 10617, Taiwan\\
	$^{10}$Physik-Department, Technische Universit\"at M\"unchen, James-Franck-Stra\ss{}e~1, 85748 Garching, Germany\\
	$^{11}$Department of Physics and Astronomy, University of California,
	Los Angeles, CA 90095, USA \\
	$^{12}$National Astronomical Observatory of Japan, 2-21-1 Osawa,
	Mitaka, Tokyo 181-8588, Japan\\
	}}

\date{Accepted XXX. Received YYY; in original form ZZZ}
\pubyear{2016}

\begin{document}
\label{firstpage}
\pagerange{\pageref{firstpage}--\pageref{lastpage}}
\maketitle

\begin{abstract}
	We present a weak gravitational lensing measurement of the external convergence along the line of sight to the quadruply lensed quasar \hequad. Using deep r-band images from Subaru-Suprime-Cam we observe galaxies down to a 3$\sigma$ limiting magnitude of $\sim 26$ mags resulting in a source galaxy density of 14 galaxies / arcmin$^2$ after redshift-based cuts. Using an inpainting technique and Multi-Scale Entropy filtering algorithm, we find that the region in close proximity to the lens has an estimated external convergence of $\kappa=-0.012^{+0.020}_{-0.013}$ and is hence marginally under-dense. We also rule out the presence of any halo with a mass greater than $M_{\rm vir}=1.6\times10^{14}h^{-1}M_\odot$ (68\% confidence limit). Our results, consistent with previous studies of this lens, confirm that the intervening mass along the line of sight to \hequad\ does not affect significantly the cosmological results inferred from the time delay measurements of that specific object.

\end{abstract}


\section{Introduction} \label{sec:intro}

In the strong regime, gravitational lensing by a galaxy-scale object can produce multiple images of the background light source. As the light rays forming the images are taking different paths to reach the observer plane, there exists a delay in the arrival time of the photons. \citet{Refsdal64} was the first to propose the use of these time delays as a tool for cosmography. Time delays are proportional to the gravitational potential of the deflector and its gradient, and to the combination of three angular diameter distances of the observer-lens-source system, dubbed the time-delay distance \citep{schneider2006, 2010ApJ...711..201S}. The latter is inversely proportional to the Hubble constant $H_0$ and has a weaker dependence on the other cosmological parameters, notably curvature and dark energy \citep[e.g.][]{2009ApJ...706...45C}. 

In practice, the values of the time delays of the lensed images are obtained by measuring the time shift between their light curves, provided they show significant variability. Although the original idea of \citet{Refsdal64} was to use lensed supernovae \citep[resolved ones found only recently, e.g.][]{2015Sci...347.1123K, 2015ApJ...811...70R, 2016arXiv161100014G}, time delays have been first measured in lensed quasars \citep[e.g.][]{1989A&A...215....1V, 1990AJ....100.1771S, 1992ApJ...384..453L}. This provided a new way to obtain independent estimates of the Hubble constant $H_0$ \citep[e.g.][]{2017MNRAS.468.2590S, 2017MNRAS.465.4914B}, complementary to and competitive with other probes such as the Cosmic Microwave Background (CMB) \citep{2016A&A...594A..13P}, Baryon Acoustic Oscillations (BAO) + CMB \citep{2017MNRAS.470.2617A}, weak lensing + BAO + Big Bang Nucleosynthesis (BBN) \citep{2017arXiv171100403D}, Cepheids \citep{2012ApJ...758...24F}, type Ia supernovae \citep{2016ApJ...826...56R}, megamasers \citep{2013ApJ...767..154R}, giant ionized $\mathrm{H}_2$ regions \citep{2017arXiv171005951F}, standard sirens \citep{2017Natur.551...85A}. 

The H0LiCOW program \citep[$H_0$ Lenses in COSMOGRAIL's Wellspring;][]{2017MNRAS.468.2590S} is devoted to the measurement of $H_0$ using lensing time delays. When completed, the goal of the first phase of the program is to measure $H_0$ to below 3.5\% precision with five gravitationally lensed quasar systems (B1608+656, RXJ1131--1231, \hequad\, WFI2033--4723 and HE1104--1805).  It mostly uses optical time delay measurements from the COSMOGRAIL collaboration (the COSmological MOnitoring of GRAvItational Lenses; e.g.  \citealt{2005IAUS..225..297C, 2011A&A...536A..53C, 2016A&A...585A..88B}), but also the radio time delays for the quadruply imaged quasar B1608+656 \citep{2002ApJ...581..823F}. In addition, H0LiCOW enables detailed study of quasar host galaxies, taking the advantage of the lensing magnifications \citep{2017MNRAS.465.4634D, 2017MNRAS.472...90D}.

In order to recover the cosmological information from an individual lens system with high precision and accuracy, the following steps are required: (i)~time-delay measurements, (ii)~modeling of the mass of the deflector using the lensing and stellar kinematic data, (iii)~modeling of the environment and the line of sight (env\&los) of the deflector (see \citealt{2016A&ARv..24...11T} for the detailed review). The H0LiCOW collaboration addresses all three steps \citep[e.g.][]{2017MNRAS.465.4914B, 2017MNRAS.465.4895W, 2017MNRAS.470.4838S, 2016arXiv160701047R}. In the present work, we focus specifically on the third step, which is needed alongside the second one in order to break the lensing degeneracies inherent to the mass-sheet transformation \citep[MST;][]{1985ApJ...289L...1F, 2013A&A...559A..37S}. 

The MST is a modification of the mass distribution which leaves all image positions, shapes and flux ratios invariant, but changes the product of Hubble constant and time delay $H_0 \Delta t$. It corresponds to the rescaling of the deflector mass by introducing a ``mass sheet'', which can be represented by a constant over- or under-density along the line of sight and around the lens. Two types of mass sheets can be distinguished: internal and external. The internal mass sheet is physically associated with the lens. The external mass sheet, on the contrary, is due to the intervening objects which do not lie in the immediate vicinity of the lens, and is thus not physically associated with it. As a consequence, the latter cannot be fully constrained by measurements of the kinematics of the main lens \citep[e.g.][]{2004astro.ph.12596K}. Modeling of the env\&los becomes mandatory. The external mass sheet can be approximated by the dimensionless surface mass density or convergence $\kappa_{ext}$ at the position of the main lens. The impact of $\kappa_{ext}$ on the Hubble constant can be expressed as
\begin{equation}
H_0^{\rm true} = (1-\kappa_{ext}) \times H_0^{\rm model},
\label{kappaext}
\end{equation}
where $H_0^{\rm model}$ is the Hubble constant obtained from the model with no account for the env\&los. Thus, neglecting the presence of an over-dense env\&los (compared with the rest of the Universe) will result in $H_0^{\rm model}$ over-estimating the true value of Hubble constant $H_0^{\rm true}$. An under-dense line of sight will have, of course, the opposite effect. For a single system, if uncorrected for, this bias can reach up to several percent \citep{2004ApJ...612..660K, 2016arXiv160105417M, 2016MNRAS.462.1405J}. The effect may not average out for the ensemble of lenses either, since due to the selection and by virtue of the lenses being typically massive ellipticals, lensed systems may be preferably observed in over-dense environments \citep{2016arXiv160508341C}.

The effects of env\&los have been evaluated in several ways. For example, \cite{2006ApJ...641..169M, 2006ApJ...642...30F, 2007AJ....134..668A, 2017arXiv171009900W} estimated $\kappa_{ext}$ by fitting analytical mass profiles to the spectroscopically confirmed groups found in the vicinity of the lens systems. \cite{2013MNRAS.432..679C} combined the halo mass model approach with the calibration to the Millennium Simulation \citep{2005Natur.435..629S}, making use of the ray-tracing calculations of convergence by \cite{2009A&A...499...31H}. 

\cite{2010ApJ...711..201S} compared galaxy number counts in the field around the lens from \cite{2011MNRAS.410.2167F} to the mock fields drawn from the Millennium Simulation. They measured $\kappa_{ext}$ in simulated fields having the same statistical properties as the real data. \cite{2013ApJ...768...39G} extended this technique by weighting the galaxy counts by distance, photometric redshift and stellar mass. 

\cite{2017JCAP...04..049B} combined the study of the environment using the halo-rendering approach, i.e. linking the galaxy stellar masses to the underlying mass distribution, with the external shear measurements of the strong lens system. Their combined approach yielded tighter constraints on the inferred external convergence compared to a halo-rendering approach only.

Finally, \cite{1997AJ....113..521F, 2009ApJ...697.1793N, 2010ApJ...711..246F} followed a different approach relying on the weak lensing effect produced by massive structures in the vicinity of the deflector. They constrained the external convergence by integrating the tangential weak gravitational shear in the area around the lens. \cite{2016arXiv160105417M} computed the weak lensing contamination preserving the full three dimensional mass distribution by using a hybrid multi-plane lensing formalism \citep{2014MNRAS.443.3631M}.

In the present work we measure the env\&los of the quadruply lensed quasar \hequad\ \citep{2000A&A...358...77W, 2002A&A...395...17W}, using the weak gravitational lensing technique. This method is direct and complementary to the env\&los study by \citet[][hereafter H0LiCOW~III]{2016arXiv160701047R}, which uses the galaxy number counts technique. The goal of our analysis is twofold. First, we estimate $\kappa_{ext}$ all the way to the redshift of the lensed quasar. Second, we test the hypothesis that the main lens may be embedded in a massive halo affecting the overall mass modeling. 

The paper is organized as follows. Section~\ref{sec:wl_formalism} contains a brief overview of weak gravitational lensing  and mass reconstruction formalism. We describe in Section~\ref{sec:data} the Subaru Suprime-Cam images used for the analysis, how we deal with masks, and the final galaxy selection. Section~\ref{sec:work} comprises the reconstruction and filtering of the convergence field, and the corrections for the lensing efficiency. Our estimates of the line-of-sight external convergence are given in Section~\ref{sec:results}. In Section~\ref{sec:sim} we test for the possible presence of a halo at a single redshift along the line of sight to the quasar, and outline the numerical simulations used to estimate the efficiency of our technique to detect such a halo, given the data. Finally, we draw our conclusions in Section~\ref{sec:summary}.

We adopt a flat $\Lambda$CDM cosmology with $\Omega_M = 0.3$,  $\Omega_{\Lambda} = 0.7$, and $H_0=70$ km/s/Mpc when producing mocks of observations. Note that the details of the cosmology chosen here have no significant effect.

\section{Weak gravitational lensing formalism}\label{sec:wl_formalism}

We summarise the formalism for weak gravitational lensing studies, from the measurement of the ellipticities of galaxies to the reconstruction of mass maps. We focus on the specific application of measuring $\kappa_{ext}$ along a given line of sight. A detailed general description of weak gravitational lensing can be found, in e.g. \citet{2001PhR...340..291B}.

\subsection{Principles of weak gravitational lensing}
\label{sec:wl}

Weak gravitational lensing manifests itself as coherent distortions of the images of distant galaxies. The effect is due to the deflection of light rays while they are propagating through an inhomogeneous gravitational field. The measure of the distortions is sensitive to the mass distribution projected along the line of sight, and depends neither on the nature nor on the physical state of the matter, hence making weak lensing an efficient mass probe. 

Weak gravitational lensing changes the apparent size, shape and magnitude of distant galaxies. The shape distortion, described by the complex shear $\gamma$, is a stretch of the image due to the 3D tidal gravitational field of the foreground mass. In the Born approximation, for sources at a single redshift $z_s$ the 3D matter distribution of the lens can be considered as an equivalent plane with a deflection potential $\psi$ \citep{schneider2006}. The complex shear field can be expressed as:
\begin{equation}
	\gamma (\pmb{\theta}, z_s) = \gamma_1 + i \gamma_2 =  (\psi_{,11} - \psi_{,22})/2 + i \psi_{,12},
	\label{gamma}
\end{equation}
where $\pmb{\theta}$ is the 2D angular position on the sky, $z_s$ is the source redshift, and $\psi$ is the 2D deflection potential described by
\begin{equation}
	\psi (\pmb{\theta}, z_s) = \frac{2}{c^2} \int_{0}^{z_s} \frac{D_{ds}}{D_{os} D_{od} } \; \Phi (\pmb{\theta}, z) \; \mathrm{d}z,
\end{equation}
where $D_{od}$, $D_{ds}$, $D_{os}$ are  the angular diameter distances between the observer and the deflector, the deflector and the source, and the observer and the source respectively, $c$ is the speed of light, and $\Phi (\pmb{\theta}, z)$ is the 3D gravitational potential of the deflector in the redshift interval $dz$.

The change in image size is caused by both anisotropic focusing of light by the tidal gravitational field, and by isotropic focusing by the local matter density. The latter is usually expressed in terms of the dimensionless effective surface mass density or convergence $\kappa$:
\begin{equation}
	\kappa (\pmb{\theta}, z_s) = \int_{0}^{z_s} \frac{\Sigma (\pmb{\theta}, z)}{\Sigma_{crit}} \; \mathrm{d}z,
\end{equation}
where $\Sigma (\pmb{\theta}, z)$ is the surface mass density in the redshift interval $dz$, and $\Sigma_{crit}$ is the critical surface mass density
\begin{equation}
	\Sigma_{crit} = \frac{c^2}{4 \pi G} \frac{D_{os}}{D_{od} D_{ds}},
\end{equation}
where $G$ is the gravitational constant. Convergence can also be expressed in terms of the deflection potential:
\begin{equation}
	\kappa = (\psi_{,11} + \psi_{,22})/2.
	\label{kappa}
\end{equation}

In practice, lensed sources are not located at a single redshift $z_s$, but rather span a redshift interval with a given distribution $p(z_s)$. This allows us to define an effective surface mass density 
\begin{equation}
\kappa = \int_{0}^{z_{max}} \kappa (\pmb{\theta}, z_s) \; p(z_s) \; \mathrm{d} z_s.
\end{equation}
where the mass is integrated along the line of sight up to a maximum redshift $z_{max}$.

\subsection{Mass reconstruction from weak gravitational lensing}
\label{sec:kappa}

In practice, the shear components $\gamma_1$ and $\gamma_2$ at some angular position $\pmb{\theta}$ can be estimated by measuring the mean ellipticity of the galaxies at that position. This process allows one to derive the shear field as described in Eq.~\eqref{gamma}. Together, Eq.~\eqref{gamma} with Eq.~\eqref{kappa} can then be inverted to reconstruct the underlying convergence map, $\kappa(\pmb{\theta})$. \citet{1993ApJ...404..441K} proposed a method to carry out this inversion in the Fourier space:
\begin{equation}
	\hat{\kappa} = P_1(\pmb{k})\hat{\gamma}_1 + P_2(\pmb{k}) \hat{\gamma}_2, 
	\label{kappa_fourier}
\end{equation}
where hat symbol denotes Fourier transform and 
\begin{equation}
	P_1(\pmb{k}) = \frac{k^2_1 - k^2_2}{k^2},
\end{equation}
\begin{equation}
	P_2(\pmb{k}) = \frac{2 k_1 k_2}{k^2},
\end{equation}
where $k^2 \equiv k^2_1 + k^2_2$, and $\pmb{k}$ is the Fourier counterparts for the angular coordinates $\pmb{\theta}$. The inverse Fourier transform of Eq.~\eqref{kappa_fourier} gives an estimate of $\kappa$. In case of an infinitely large field of view, this estimate is correct up to an overall additive constant, as constant surface mass density does not cause any shear and is thus unconstrained by $\gamma$.

The mass map $\kappa$ derived in this way is unreliable due to noise caused by the limited number density of galaxies with measurable ellipticities, the finite field of view and the loss of some areas due to bright or large foreground objects. Advanced filtering techniques must be used, in particular when dealing with small fluctuations in the convergence field. In this work we apply both the original method by \citet{1993ApJ...404..441K} using smoothing with a Gaussian kernel, and a more advanced technique based on wavelets (see Section~\ref{sec:mrlens}). Our goal is to measure $\kappa$ in the best possible way at the position of \hequad\ on the plane of the sky, i.e. $\kappa_{ext}$ in Eq.~\eqref{kappaext}.

\subsection{E and B modes of the shear field}
\label{sec:ebmodes}

The validity of the reconstructed mass maps can be tested by decomposing the shear field into an ``electric'' E-mode and a curl or ``magnetic'' B-mode map. As the shear field arises from a scalar gravitational potential, in the absence of lens-lens coupling or higher order effects, only E-modes should be present in the reconstructed mass maps. Residual systematics introduced by sky subtraction or correction for the PSF can generate both E- and B-modes (e.g. \citealt{2004ApJ...613L...1V}). Detection of significant B-modes may therefore indicate the presence of such systematics. 

The E-mode maps can be transformed into the B-mode maps by rotating the shear by $45^{\circ}$: $\gamma_1 \rightarrow -\gamma_2; \; \gamma_2 \rightarrow \gamma_1 $. As a result, while Eq.~\ref{kappa_fourier} gives the E-mode part of the convergence map, an estimator for the B-mode convergence field is:
\begin{equation}
	\hat{\kappa}_B = P_2(\pmb{k})\hat{\gamma}_1 - P_1(\pmb{k}) \hat{\gamma}_2.
\end{equation}

\section{Observations}
\label{sec:data}

\begin{figure*}
	\centering
	\captionsetup[subfigure]{labelformat=empty}
	\subfloat{
		\includegraphics[width=1.018\columnwidth]{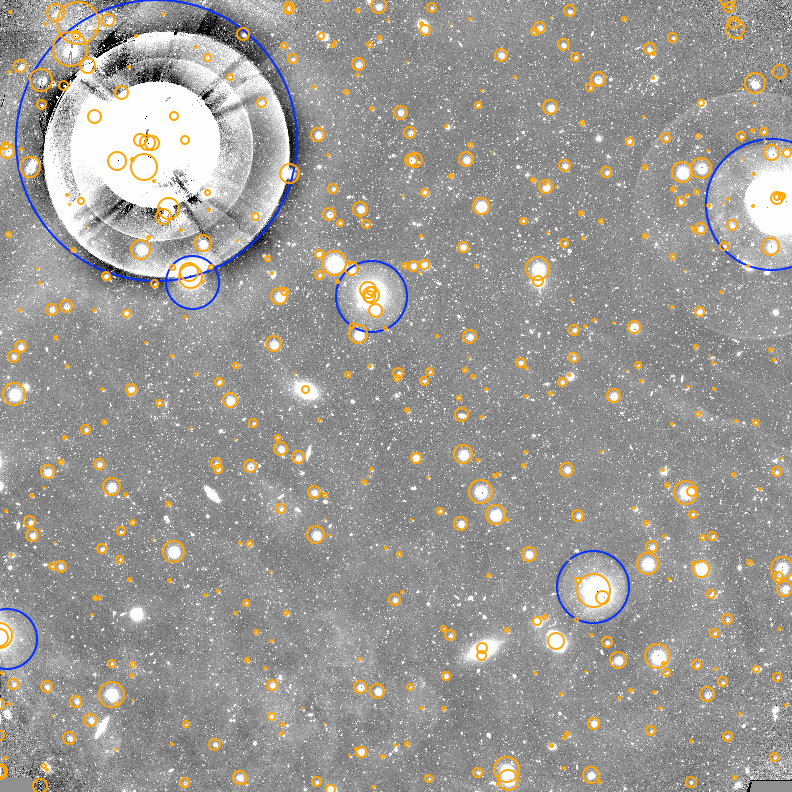}
	}
	\subfloat{
		\includegraphics[width=1.018\columnwidth]{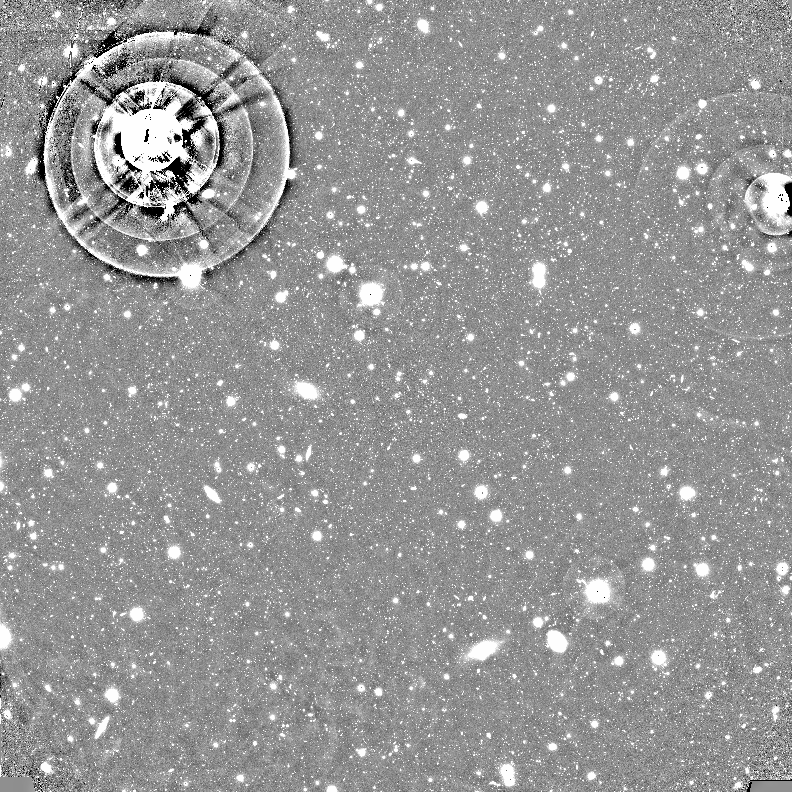}
	}\\
	\vspace*{-0.2cm}	
	\subfloat{
		\includegraphics[width=0.67\columnwidth]{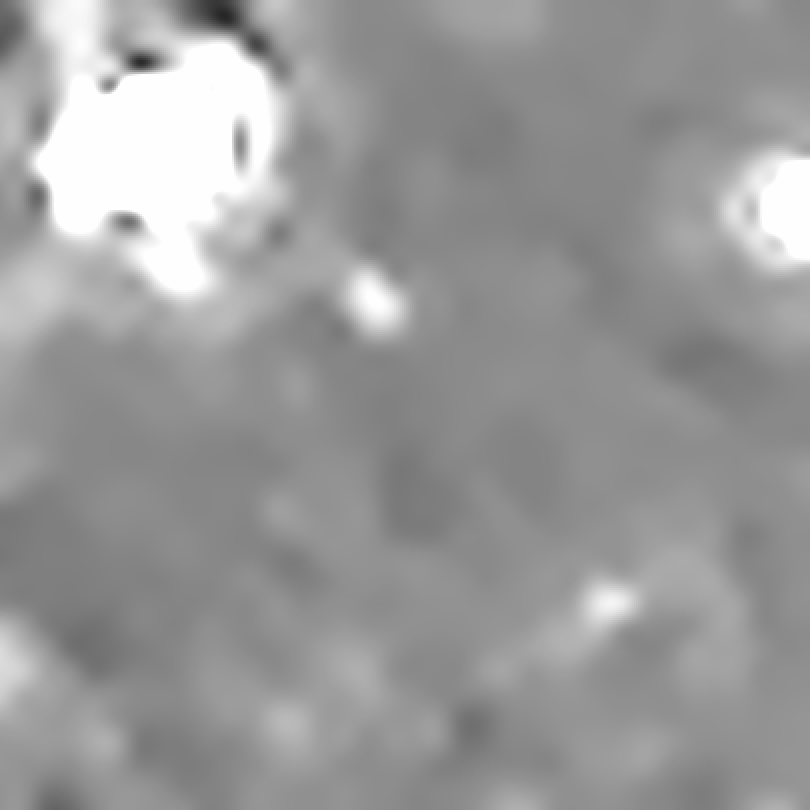}
	}
	\subfloat{
		\includegraphics[width=0.67\columnwidth]{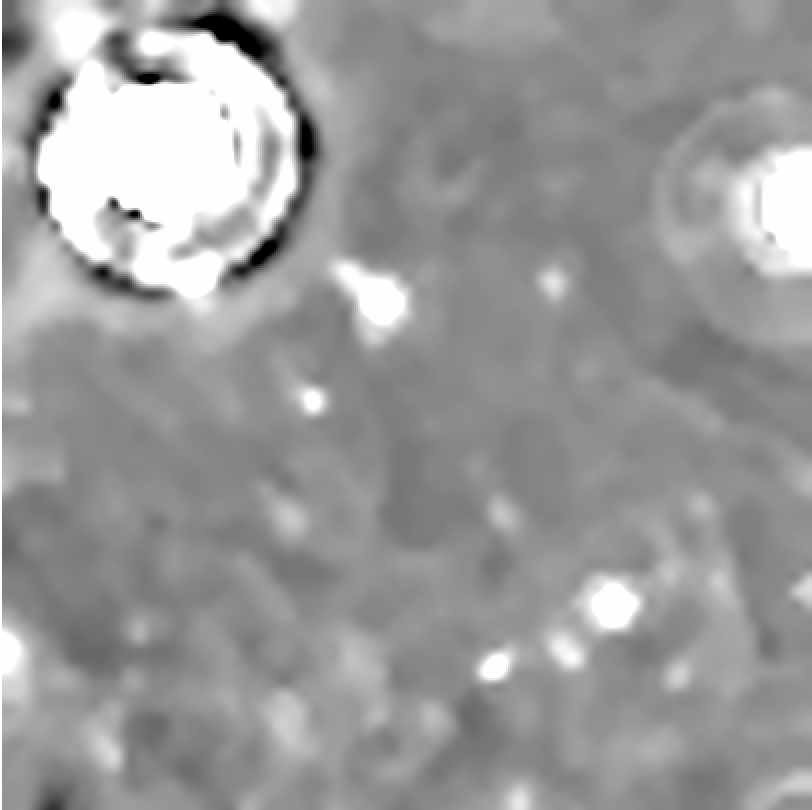}
	}
	\subfloat{
		\includegraphics[width=0.67\columnwidth]{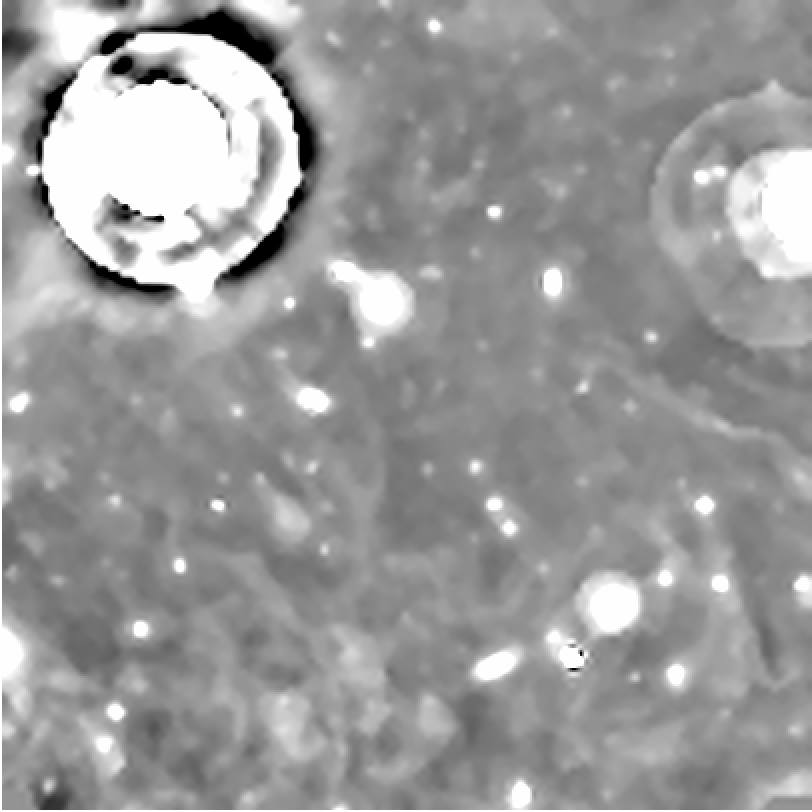}
	}
	
	\caption{The $0.5^{\circ} \times 0.5^{\circ}$ field around \hequad, with North up and East on the left. The top left panel shows the original image. The blue circles indicate the areas masked due to saturated stars, and the orange circles show the areas masked using our automated $\tt{SExtractor}$ procedure. The top-right panel shows the final background-subtracted image used for the weak lensing analysis. We show for reference different background models depending on the last scales adopted for the $mr\_background$ algorithm (see the text): bottom left panel -- 6th scale, bottom central panel -- 7th scale, bottom right panel -- 8th scale. For our analysis we adopt the 7th scale.}
	\label{fig:0435}
\end{figure*}

The imaging data for the quadruple quasar \hequad\ (RA(2000) = 04h 38min 14.9 sec; DEC(2000) = -12$^{\circ}$ 17\arcmin\ 14\farcs4) was obtained on the 1st of March 2014 with the Suprime-Cam instrument mounted on the 8.2m Subaru telescope\footnote{Program ID: S14A-TE083; PI: C.~D.~Fassnacht}. The strongly lensed quasar is located at the redshift of $z_s = 1.693$ \citep{2012A&A...544A..62S} and the lens galaxy is at the redshift of $z_l = 0.4546$ \citep{2005AJ....129.2531M, 2006A&A...451..759E}. The reduction and calibration of the data are described in detail in H0LiCOW~III. Additionally, they make use of the multi-band optical and near-IR imaging data to calculate the photometric redshifts for the galaxies. In the present work we use the deep $r$-band data for the weak lensing analysis. The $r$-band image has a 3$\sigma$ limiting magnitude of $r=25.94 \pm 0.28$, a seeing of 0.7\arcsec, and a mean airmass of $a=1.7$. The image is a combination of 16 exposures of 300 sec each, with a pixel scale of $0.200 \arcsec$. The resulting useful field of view is 34\arcmin $\times$ 27\arcmin, as shown in Fig.~\ref{fig:0435}.

\subsection{Sky subtraction and masking}\label{sec:sky}

Subtracting the sky background is a critical step of the weak lensing analysis. In the case of our data, the sky background includes imperfect illumination of the field as well as Galactic cirrus in the Southern part of the field. We subtract this foreground light contamination using the $\tt{mr\_background}$ algorithm, which is part of the Multiresolution Analysis Software \footnote{http://www.multiresolutions.com/mr/}. In this multi-scale analysis, the background is considered to be the last scale of a pyramidal median transform \citep{1996PASP..108..446S} of the image. The number of scales is automatically calculated so that the size of the last scale is lower than or equal to $N \times N$ pixels, where $N$ is the size of a user-specified box.

We try different last scales, i.e. the 6th, the 7th and the 8th scales. For a $9000\times9000$ pix$^2$ image this corresponds to $N \approx 140, 70, 35$ pixels, respectively. We find that the 7th scale is the one that models best the spatial structures in the sky background down to the noise level, without affecting the flux of the small objects. While the 6th scale under-fits the background data, i.e. leads to an overly smoothed background image, the 8th scale picks up too many fine details belonging to stars and galaxies, i.e. it over-fits the data (see the bottom row of Fig.~\ref{fig:0435}). The result of this background subtraction process is shown in the upper right panel of Fig.~\ref{fig:0435}. We also try to remove the sky using the $\tt{SExtractor}$ software\footnote{http://www.astromatic.net/software/sextractor} \citep{1996A&AS..117..393B}, but the multi-scale approach of the $\tt{mr\_background}$ provides cleaner subtraction.

We apply the $\tt{SExtractor}$ software for the detection and primary measurements of galaxy and star properties. All the stars flagged as saturated by $\tt{SExtractor}$ are masked out (see Fig.~\ref{fig:0435}). We apply circular masks with radius $r = 2\times \mathtt{FWHM\_IMAGE}$ , where $\mathtt{FWHM\_IMAGE}$ is the full width at half maximum. We also manually mask 6 very bright stars with luminous halos that extend up to 1\arcmin\ in radius and that are not detected by $\tt{SExtractor}$. The radius of the masks corresponds to roughly $5\sigma$ of the Gaussian light profile of these stars, so that the mask contains almost all the flux of the saturated object.

\subsection{Galaxy selection}
\label{sec:stargals}

\begin{figure}
	\centering
	\includegraphics[width=0.95\columnwidth]{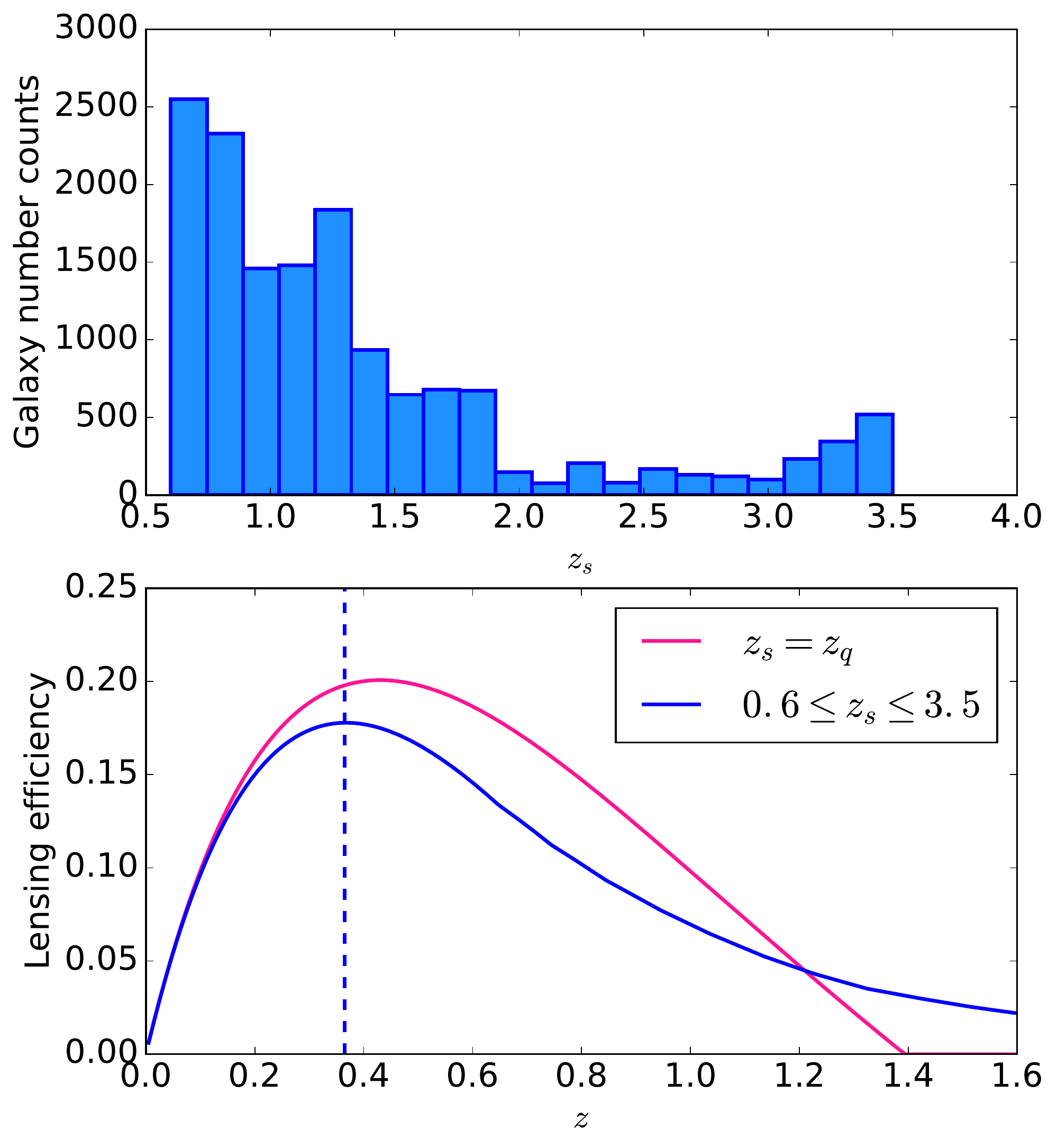}
	\caption{The redshift distribution of the galaxies used in the weak lensing analysis is shown on the top panel. The lower panel shows the cumulated lensing efficiency kernel computed for the selected source galaxies with $0.6 \leq z_s \leq 3.5$ (blue), as compared to the lensing efficiency kernel for the galaxies at the redshift of the lensed quasar, $z_q$ (pink). The dashed blue line shows the lens redshift, for which the cumulated lensing efficiency is maximal.}
	\label{fig:kernel}
\end{figure}

We use all the Subaru Supreme-Cam $ugri$-band images for the star-galaxy classification and photometric redshift estimates, adopting the same techniques as in H0LiCOW~III but extending them to fainter magnitudes. The final catalogs cover the full extent of the Subaru $r$-band image. We keep the galaxies with photometric redshift uncertainties $\sigma_z< 0.35 \times (1 + z)$, where $\sigma_z$ is the 68\%-level errors.

In order to estimate the impact of the env\&los on the determination of $H_0$, it is necessary to measure the surface matter density $\kappa_{ext}$, projected all the way to the redshift of the quasar. Ideally, this is achieved by selecting a single plane of source galaxies at the redshift of the \hequad\ quasar $z_q = 1.693$. In practice, however, such a drastic selection leads to a galaxy number density too low for adequate weak lensing measurements. To overcome this, we select the source galaxies so that their cumulated lensing efficiency kernel is as close as possible to the kernel for a single plane at $z_q$, while maintaining a reasonable galaxy number density. We bin the redshifts of the source galaxies and compute the cumulated kernel as
\begin{equation}
\mathcal{G}(z_l) = \frac{\sum\limits_{i} \; g(z_l, z_s^i) \; n(z_s^i)}{\sum\limits_{i} n(z_s^i)},
\end{equation}
where $n(z_s^i)$ is the number of source galaxies per redshift bin~$i$, $z_s^i$ is the central source redshift of the bin, $z_l$ is the redshift of the lens, and $g(z_l, z_s^i)$ is the lensing efficiency kernel calculated for $z_s^i$, i.e.
\begin{equation}
g(z_l, z_s^i) = \frac{4 \pi G}{c^2} \frac{D_{ol} D_{ls}}{D_{os}} \; H(z_l-z_s^i) = \frac{H(z_l-z_s^i)}{\Sigma_{crit}}.
\end{equation}
$H(z_l-z_s^i)$ is the Heaviside step function, which accounts for the fact that the sources in front of the structures in a given bin are not lensed. The summation is done over all redshift bins of source galaxies. 

We find that selecting the source galaxies in the range $0.6 \leq z_s \leq 3.5$ (see upper panel of Fig.~\ref{fig:kernel}) minimizes the difference between the optimal lensing kernel at $z_s = z_q$ and the resulting cumulated one. This leaves us with a total of 12569 galaxies useful for the weak lensing analysis, or 14~gals/arcmin$^2$. The two kernels are shown in Fig.~\ref{fig:kernel}. While the cumulated lensing efficiency kernel is close to the ideal kernel, it does not match it perfectly. We therefore correct the effect of the mismatch between the ideal kernel and the one imposed on us by the data using N-body simulations. This last step is described in detail in Sect.~\ref{sec:efficiency_correction}. 

Note that the uncertainties on the photometric redshifts might influence the cumulated lensing efficiency kernel. This can be tested in the following way. Instead of taking the peak of the photometric redshift estimate, for each galaxy we allocate the redshift drawn from its own redshift distribution as computed from the photo-z procedure. This changes the overall histogram shown in the upper panel of Fig.~\ref{fig:kernel}. We then calculate the new cumulated efficiency kernel. We repeat the procedure 1000 times, and find that the spread between the resulting curves is negligible compared to the difference between the cumulated and the ideal kernels.

\section{Mass mapping of the line of sight to \hequad}
\label{sec:work}

We produce the galaxy shape catalog using the $\tt{KSB+}$ software \citep{2006MNRAS.368.1323H}, which accounts for the point spread function (PSF) and its spatial variations in the data. $\tt{KSB+}$ is a refined version of an algorithm initially developed by \cite{1995ApJ...449..460K}. It approximates the PSF as a small but highly anisotropic distortion convolved with a large circularly symmetric seeing disk. $\tt{KSB+}$ parametrizes objects according to their weighted quadruple moments and provides directly the shear estimator $\gamma$ for each galaxy. 

From the ellipticity catalog we reconstruct the shear field, accounting for edge effects and the missing data due to masking of the bright stars. We then reconstruct the convergence field using two different techniques to minimize the impact of shot noise. Finally, we correct for the difference in the lensing efficiency kernels for our selection of galaxies and for that of a screen of background galaxies at $z=z_q$.

\subsection{Reconstruction of the shear field}\label{sec:fastlens}

In order to reconstruct the convergence mass map, we make the Fourier transform of the shear field, following the equations in Section~\ref{sec:kappa}. This must be done with care, as the masks and field edges produce high frequency signals that are aliased in the Fourier domain. To alleviate this effect, we use the $\tt{FASTLens}$\footnote{http://www.cosmostat.org/software/fastlens/} software \citep{2009MNRAS.395.1265P}.

$\tt{FASTLens}$ implements an inpainting algorithm, which is used to fill in the gaps in the data by extrapolating the existing information. The inpainting technique is successfully used in various astrophysical areas, e.g. Cosmic Microwave Background \citep{2010A&A...519A...4P, 2012A&A...544A..27P, 2013A&A...550A..15S} and asteroseismology \citep{2014A&A...568A..10G, 2015A&A...574A..18P}. The $\tt{FASTLens}$ algorithm is set up in the Bayesian framework, using sparsity of the solution as a prior \citep{elad2005}. It assumes that there exists a transform dictionary, where the complete data is more sparse than the incomplete data. In the weak lensing case, a well suited dictionary proves out to be the Discrete Cosine Transform \citep[DCT;][]{2009MNRAS.395.1265P}. In the DCT domain the weak lensing signal becomes sparse, meaning that the majority of the coefficients into which the signal is decomposed turn out to be negligible. The masks introduce additional coefficients not related to the original data, that can be removed by thresholding. The solution is obtained though an iterative process with exponentially decreasing thresholds, where the number of iterations is fixed empirically \citep{2009MNRAS.395.1265P, 2015A&A...574A..18P}. 

When using the $\tt{FASTLens}$, we bin our shear map so that each spatial resolution element contains at least 1 galaxy. For the present data this implies the reduction of the original image size from $9000\times9000$ pixels$^2$ to $256\times256$ pixels$^2$, each pixel being 0.1\arcmin\ on-a-side. We perform various tests to estimate the number of iterations, that we finally fix to 300. Increasing this number does not change the result significantly. Note that, for the rest of the study, we are not using the inpainted data inside the masks. The technique is only used to avoid the artifacts produced by the aliased frequencies coming from the masks and field edges, that contaminate the signal.

\subsection{Reconstruction of the convergence map and noise filtering}
\label{sec:mrlens}

We use the standard Kaiser\&Squires technique to convert the shear field into a convergence field (see Section~\ref{sec:kappa}). The original method by \cite{1993ApJ...404..441K} uses a Gaussian convolution kernel with varying aperture size, $\theta_G$, to filter the data. This filtering technique is linear, easy to implement and is widely used in the field of weak lensing \citep[e.g.][]{2015PhRvD..92b2006V}. It has significant drawbacks, though. When large aperture sizes are used, smaller features in the map might get smoothed out, resulting in a loss of resolution. On the other hand, small $\theta_G$ values lead to larger reconstruction errors \citep{2006A&A...451.1139S}. Finally, the choice of $\theta_G$ itself is somewhat arbitrary. 

An alternative approach to Gaussian filtering is multi-scale entropy filtering \citep[MSE;][]{2006aida.book.....S}, successfully used in different applications of weak lensing \citep[e.g.][]{2009A&A...505..969P, 2016A&A...593A..88L}. MSE filtering is a non-linear Bayesian filtering technique which uses a MSE prior. 

The data is decomposed into multiple scales using the ``\`a trous'' wavelet transform \citep{1989wtfm.conf..286H}. This transform ensures the sparsity of the lensing signal at all spatial scales. Importantly, the noise in this dictionary is non-sparse, while the lensing signal is. As a consequence, most of the lensing information is described by few highly significant coefficients, while the noise is spread over many non significant coefficients. These are removed using the false discovery rate \citep[FDR;][]{fdr, 2001AJ....122.3492M}, which adapts the selection threshold based on the desired fraction of false detections over the total number of detections.

The MSE prior is constructed using all the non-significant coefficients in each of the wavelet scales. The entropy is calculated by modeling the noise in the data \citep{2001A&A...368..730S}, which produces good results for the analysis of piecewise smooth images and is thus well adapted for the mass reconstruction \citep{2006A&A...451.1139S}. 

We apply both Gaussian and the MSE filtering techniques to the images of \hequad. 
For the Gaussian filtering we adopt two different kernel apertures: $\theta_G = 0.5'$ and $\theta_G = 1'$, which allow us to achieve a fairly good spatial resolution, while still preserving the Gaussian properties of the noise \citep[e.g.][]{2000MNRAS.313..524V}. For the MSE filtering, we use the MSE algorithm implemented in the $\tt{MRLens}$\footnote{http://www.cosmostat.org/software/mrlens/} software \citep{2006A&A...451.1139S}. We decompose the original image into 6 wavelet scales, and filter the first 5 scales, starting from the highest spatial frequencies. We specify the fraction of false detections to be  $\alpha = 0.01$ (1\%) for the first scale, which roughly corresponds to a $3 \sigma$ thresholding \citep{2001AJ....122.3492M}, where $\sigma$ is the noise standard deviation. This fraction is gradually decreased by a factor of 2 for every subsequent scale.

\subsection{Correcting for the lensing efficiency}
\label{sec:efficiency_correction}

\begin{figure*}
	\centering
	\captionsetup[subfigure]{labelformat=empty}
	\subfloat{
		\includegraphics[width=1.0\textwidth]{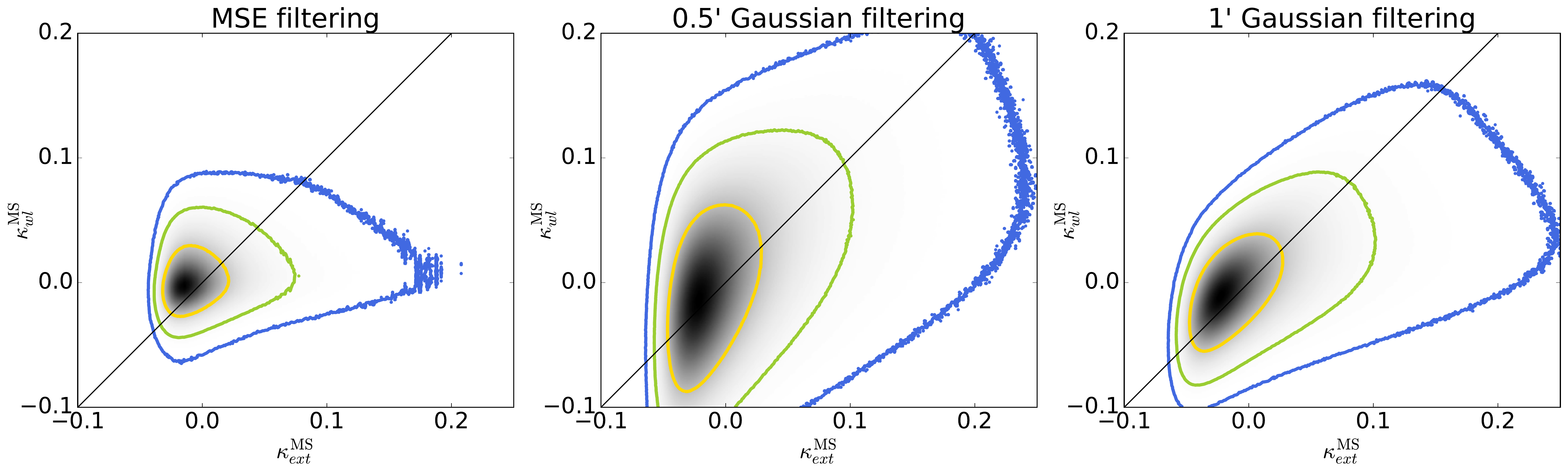}}
	\caption{Joint distributions for the convergence inferred from the Millennium simulation for the sources following our observed redshift distribution, i.e. $\kappa_{wl}^{\mathrm{MS}}$, and for the sources lying in a single redshift plane, i.e. $\kappa_{ext}^{\mathrm{MS}}$. The left panel corresponds to the MSE filtering of the data, the middle panel to the Gaussian filtering with $\theta_G = 0.5'$, and the right panel to the Gaussian filtering with $\theta_G = 1'$. The yellow, green and blue contours show the $1\sigma$, $2\sigma$ and $3\sigma$ regions respectively. The black line indicates the perfect correlation.}
	\label{fig:kernel_distribution}
\end{figure*}

Our selection of galaxies, which maximizes the number density of measurable sources, yields an effective lensing efficiency kernel which is somewhat different from the ideal one where all source galaxies lie at the redshift of the quasar $z_s = z_q$. What we measure from our data $\mathrm{\bf{d}}$ is thus $P(\kappa_{wl} | \mathrm{\bf{d}})$. To account for possible resulting mis-estimation of the external convergence, we calibrate our measurement using ray-tracing through the Millennium Simulation. In this way we estimate $ P(\kappa_{ext} | \mathrm{\bf{d}})$, which is the final result of this analysis.

The Millennium Simulation \citep[MS,][]{2005Natur.435..629S} is a large high-resolution cosmological simulation, run with GADGET-2 code \citep{2005MNRAS.364.1105S} assuming the $\Lambda$CDM model of hierarchical structure formation. It follows the evolution of $N = 2160^3$ dark matter particles with masses of $8.6 \times 10^8 h^{-1} M_{\odot}$ from redshift $z = 127$ to $z = 0$ in a cubic region with comoving side length $L = 500 h^{-1} \: \mathrm{Mpc}$. The resulting MS dark matter halos are then populated with galaxies using the semi-analytic galaxy models by \citet{2007MNRAS.375....2D}.

\citet{2009AA...499...31H} estimated the lensing effect produced by the dark matter structures of the MS using the ray-tracing algorithm based on the multiple-lens-plane approximation. This multiple-lens-plane approximation allows us to calculate lensing observables by projecting the continuous mass distribution on discrete lens planes with an accuracy of a few percent. \citet{2008MNRAS.386.1845H} took into account the additional effects of luminous matter from galaxies with stellar masses $\geq 10^9 h^{-1} M_{\odot}$, which were taken from the catalog by \citet{2007MNRAS.375....2D}.

We select 1024 fields from the MS with the field of view of $0.5^{\circ} \times 0.5^{\circ}$, which we populate with our source galaxies, preserving their position in the 3D space $(RA, DEC, z_{phot})$. Following \citet{2008MNRAS.386.1845H, 2009AA...499...31H}, we perform the multiple-lens-plane ray-tracing to calculate the shear for all source galaxies. For each of 1024 fields we generate 1000 noisy realisations by adding the shape noise, which is drawn from a normal distribution with $\sigma_{\gamma} = 0.25$, as estimated from the data. We then apply our mass map reconstruction and noise filtering methods on these simulated fields to measure the weak lensing convergence maps $\kappa_{wl}^{\mathrm{MS}}$. 

Using the same ray-tracing technique we calculate the shear for the source galaxies at $(RA, DEC, z_{q})$, i.e. maintaining their 2D position, but placing them all at the redshift of the quasar $z_q$. After performing the reconstruction, we produce $\kappa_{ext}^{\mathrm{MS}}$ maps for the case of the ideal kernel, i.e. with all source galaxies lying at the same redshift. 

 The simulations we carry out provide $P(\kappa_{ext}^{MS} | \kappa_{wl}^{MS})$ for all three filtering techniques, as shown in Fig.~\ref{fig:kernel_distribution}. We define $\kappa_{ext}$ as the corrected external convergence that has all the sources located at the quasar redshift. We can obtain $\kappa_{ext}$ from the $\kappa_{wl}$ that we measure from the data using our pipeline by identifying $\kappa_{wl}^{MS}$ with $\kappa_{wl}$ and $\kappa_{ext}^{MS}$ with $\kappa_{ext}$, such that
 \begin{equation}
  P(\kappa_{ext} | \kappa_{wl}) = P(\kappa_{ext}^{MS} | \kappa_{wl}^{MS}).
 \end{equation}  
 The probability density function (PDF) of $\kappa_{ext}$ given the available data $\mathrm{\bf{d}}$, i.e. $P(\kappa_{ext} | \mathrm{\bf{d}})$, is then 
 \begin{equation}
 P(\kappa_{ext} | \mathrm{\bf{d}}) = \int \mathrm{d} \kappa_{wl} \,\, P(\kappa_{ext} | \kappa_{wl}) \times P(\kappa_{wl} | \mathrm{\bf{d}}).
 \end{equation}

\section{Line of sight convergence}
\label{sec:results}

\begin{figure}
	\centering
	\captionsetup[subfigure]{labelformat=empty}
	\subfloat{
		\includegraphics[width=0.48\textwidth]{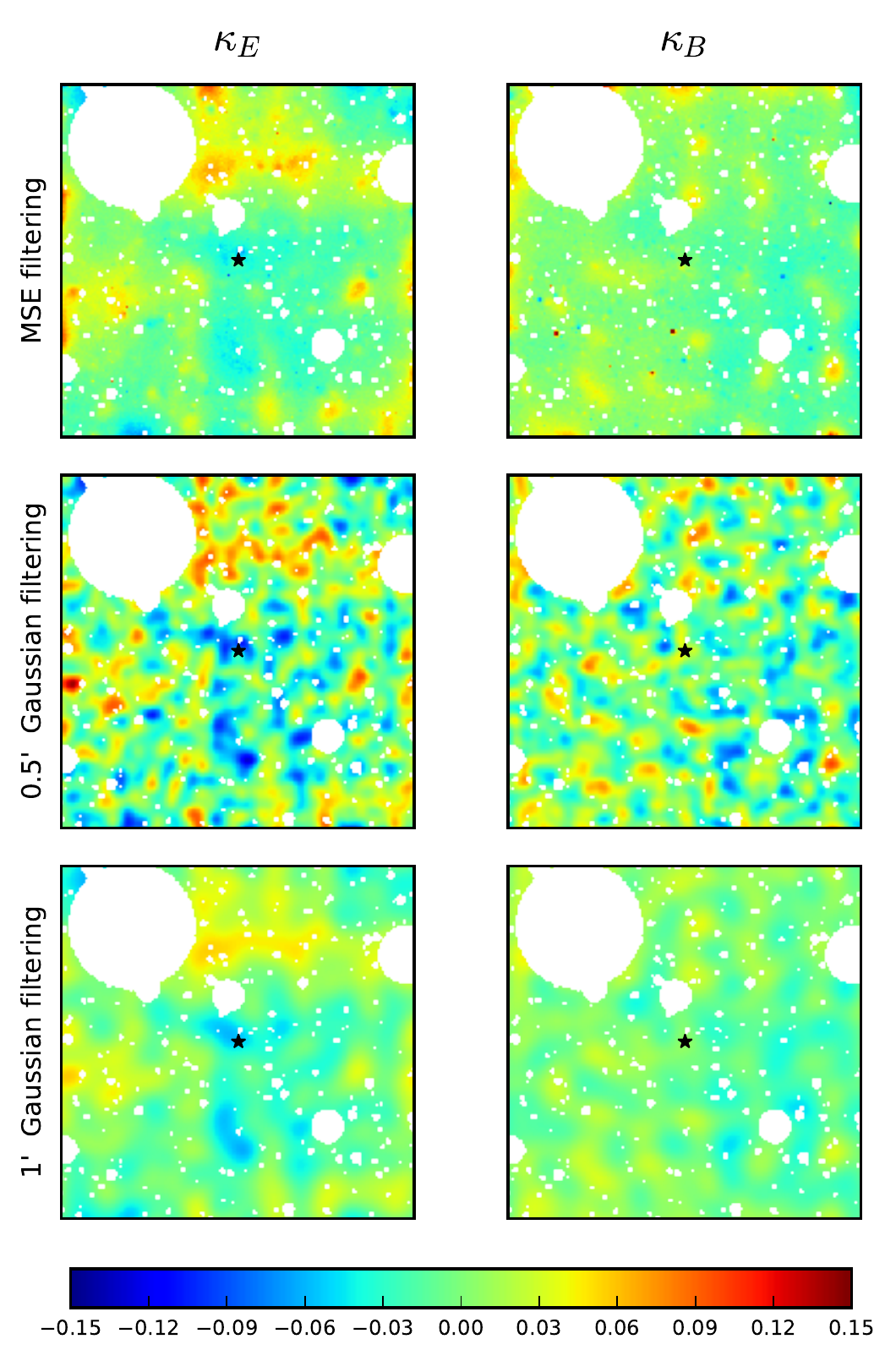}}
	\caption{Convergence maps for the $0.5^{\circ} \times 0.5^{\circ}$ field around \hequad, indicated by a star in the center. The left column shows the E-modes and the right column the B-modes. The upper row corresponds to the MSE filtering of the data, the middle row to the Gaussian filtering with $\theta_G = 0.5'$, and the bottom row to the Gaussian filtering with $\theta_G = 1'$. All maps have pixel scale of $0.1'$. The white regions correspond to masked stars and bright foreground objects.}
	\label{fig:kappa}
\end{figure}

\begin{figure*}
	\centering
	\captionsetup[subfigure]{labelformat=empty}
	\subfloat{
		\includegraphics[width=0.8\textwidth]{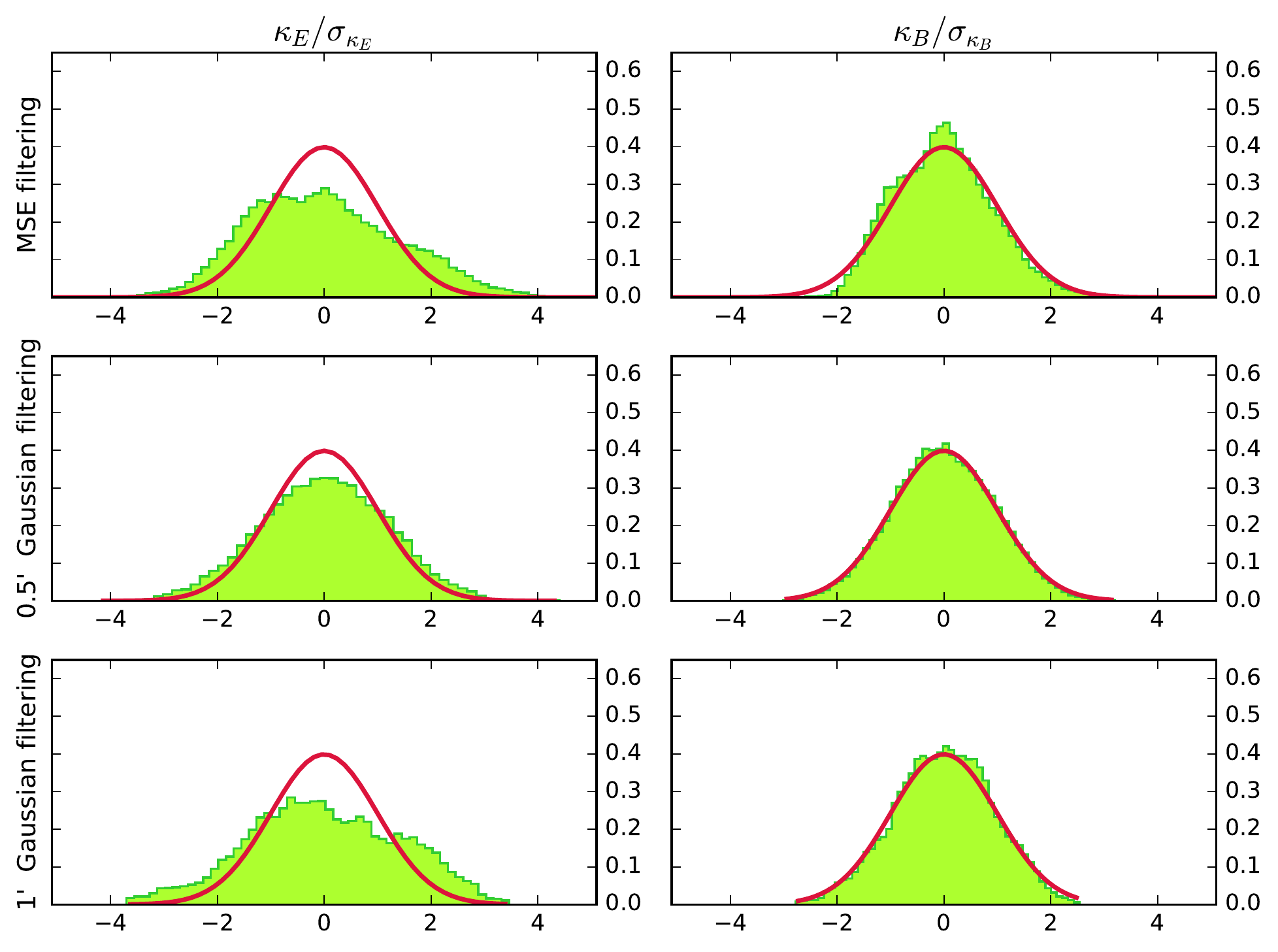}}
	\caption{Normalized S/N distributions of convergence maps with E-modes shown on the left and the B-modes on the right. From top to bottom are shown the MSE filtering, and the Gaussian filtering with $\theta_G = 0.5'$, and $\theta_G = 1'$. In each panel, the solid line shows a normal distribution centered at zero with a unit standard deviation. The B-modes are compatible with a normal distribution for all filtering techniques, while E-modes deviate significantly, as expected in the presence of a weak lensing signal.}
	\label{fig:pool_plot}
\end{figure*}

Figure~\ref{fig:kappa} shows our measured convergence maps for \hequad, produced using the data described in Section~\ref{sec:data} and the analysis pipeline detailed in Sect.~\ref{sec:work}. The maps are given for each of the three filtering techniques and for both E-modes and B-modes, prior to the correction described in Sect.~\ref{sec:efficiency_correction}. As lensing does not produce B-modes, the corresponding map should not show any significant B-mode signal. This is the case for our data according to the right-column of Fig.~\ref{fig:kappa}. 

To quantify this further, we estimate the statistical uncertainties by rotating the galaxies by a random angle, preserving their initial shapes and positions. We generate 1000 such shear fields, which we analyse with our pipeline to produce 1000 corresponding $\kappa_E$ and $\kappa_B$ maps for each filtering technique. The standard deviation between these 1000 maps provides a corresponding noise map, which contains the galaxy shape noise and the measurement error. We then divide the original maps by the noise maps to estimate the S/N of the structures. 

From these S/N maps, we generate the S/N distributions for the E-modes and the B-modes. In the absence of the signal, the B-modes should be consistent with a normal distribution centered at zero with a standard deviation equal to one. The S/N distribution of the E-modes should, on the contrary, have a standard deviation greater that one \citep[e.g.][]{2014ApJ...786...93U}. Our results are displayed in Fig.~\ref{fig:pool_plot}, where indeed the B-modes agree with Gaussian distribution for all three filtering techniques. The E-modes deviate from Gaussian distribution, which is indicative of a lensing signal. 

\begin{figure*}
	\centering
	\captionsetup[subfigure]{labelformat=empty}
	\subfloat{
		\includegraphics[width=0.9\textwidth]{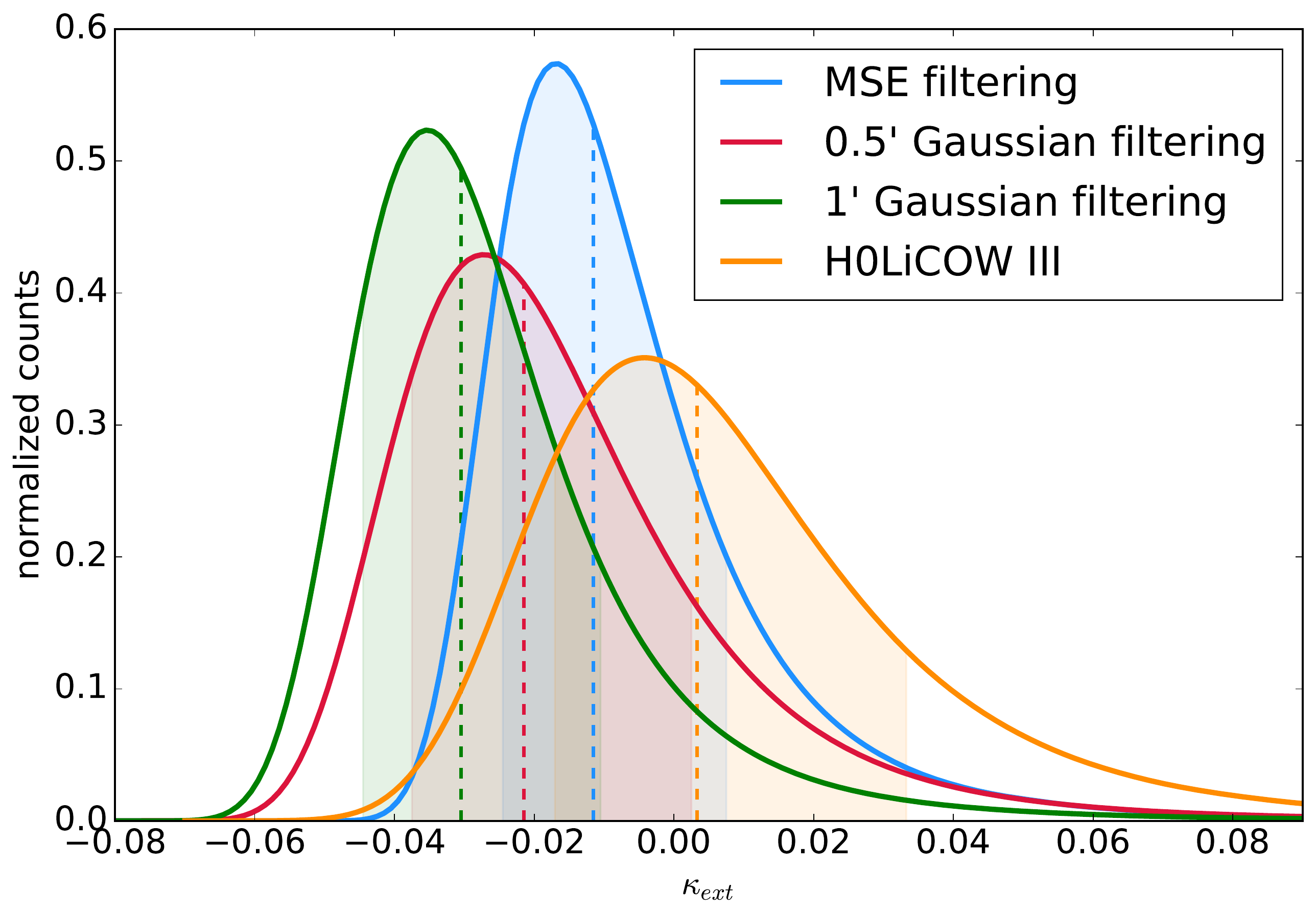}}
	\caption{PDFs of external convergence for the \hequad\ field. The blue distribution corresponds to the MSE filtering of the data, red to the Gaussian filtering with $\theta_G = 0.5'$, and green to the Gaussian filtering with $\theta_G = 1'$. The orange distribution shows the result from H0LiCOW~III. Dashed lines show the values at the $50\%$ percentile. Shaded regions indicate the $1\sigma$ regions.}
	\label{fig:final_result}
\end{figure*}

Finally, to estimate the external convergence along the line of sight to the \hequad\ system, we measure the convergence inside the central pixel of our mass maps. We use the 1000 noise realisations to construct $P(\kappa_{wl} | \mathrm{\bf{d}})$ for each noise filtering method, which we center on the values obtained from the corresponding signal maps. We then correct the PDFs for the difference in the lensing efficiency kernel, by weighting them with the joint distributions obtained from the Millennium simulation, as described in Section~\ref{sec:efficiency_correction}, which yields $P(\kappa_{ext} | \mathrm{\bf{d}})$. This correction also accounts for the smaller scales, possibly omitted in our analysis due to the noise filtering. Our final $P(\kappa_{ext} | \mathrm{\bf{d}})$ are displayed in Fig.~\ref{fig:final_result}, together with the results from H0LiCOW~III based on weighted galaxy counts. The values of $\kappa_{ext}$ at the position of \hequad\ and associated error bars are given in Table~\ref{tab:kappa_values}.

\begin{table}
	\begin{center}
		\begin{tabular}{ c c c c c}
			\hline
			Filtering technique & $\kappa_{ext}$ & $\sigma_{\kappa}^{-}$ &  $\sigma_{\kappa}^{+}$ & F \\
			\hline\hline
			MSE & -0.012 & 0.013 & 0.020 & 5.1 \\
			G$0.5'$ & -0.022 & 0.016 & 0.025  & 3.7 \\
			G$1'$ & -0.031 & 0.014 & 0.040 & 2.7 \\
			\hline
			H0LiCOW~III & 0.003 & 0.020 & 0.030 & $-$\\
			\hline
		\end{tabular}
	\end{center}
	\caption{External convergence estimates for the \hequad\ field using three different noise filtering techniques: MSE filtering, Gaussian filtering with $\theta_G = 0.5'$ and $\theta_G = 1'$. The values are given at the position of the quasar. For the comparison, we also give the values from H0LiCOW~III. $\kappa_{ext}$ show median value, $\sigma_{\kappa}^{-}$ and $\sigma_{\kappa}^{+}$ correspond to deviation from 16th and 84th quantiles respectively. For each of the filtering techniques we indicate the Bayes Factor calculated with respect to the H0LiCOW~III result. (see text).}
	\label{tab:kappa_values}
\end{table}

\subsection{Comparison with the results from H0LiCOW~III}

We now investigate the consistency between the weak lensing ($\kappa^\mathrm{wl}_{ext} \equiv \kappa^\mathrm{wl}$, this paper) and weighted number counts ($\kappa^\mathrm{nc}_{ext} \equiv \kappa^\mathrm{nc}$, H0LiCOW~III) techniques. To do so, we adopt the Bayesian formalism proposed in \citet{2006PhRvD..73f7302M} and we test two hypotheses: 

\begin{enumerate}
	
\item $H^\mathrm{global}$: both results can be consistently explained within one set of cosmological parameters, describing the same field and environment. 

\item $H^\mathrm{ind}$: there are some unaccounted systematic errors leading to an offset, which can be parametrized with a second independent set of cosmological parameters. In this case, two sets of parameters are needed to account for the two results separately, as they are describing two different environments.

\end{enumerate} 

To infer which hypothesis is favored by the data, we calculate the Bayes factor $F$, given by
\begin{equation}
F = \frac{ P(\kappa^\mathrm{wl}, \kappa^\mathrm{nc} | H^\mathrm{global}) }{ P(\kappa^\mathrm{wl}, \kappa^\mathrm{nc} | H^\mathrm{ind}) }.
\end{equation}
In our case, this reduces to
\begin{equation}
F = \frac{ \langle L^\mathrm{wl} \, L^\mathrm{nc} \rangle }{ \langle L^\mathrm{wl} \rangle  \langle L^\mathrm{nc} \rangle },
\end{equation}

where $L^\mathrm{wl}$ and $L^\mathrm{nc}$ are respectively the likelihoods of an external convergence obtained from the weak lensing and from the weighted number count methods (see Appendix in \citet{2013ApJ...766...70S} for the derivation of the formula). If $F>1$, the data favors hypothesis $H^\mathrm{global}$ describing the same environment. 

The values for the Bayes factor $F$ for three filtering techniques are given in Table~\ref{tab:kappa_values}. For reference, two one-dimensional Gaussian likelihoods have a Bayes factor of $F = 1$ if they overlap within $2\sigma$'s, and $F\sim 3.6$ if the two distributions overlap within $1\sigma$. As in our case all Bayes factors are noticeably larger than 1, we conclude that the convergence estimates by the weak lensing technique and by the weighted galaxy number count technique provide consistent results. We also see that the MSE noise filtering technique that preserves the best the smaller scales is more in agreement with the weighted galaxy number count technique. The Gaussian filtering with $\theta_G = 1$\arcmin\ is the least consistent, probably as it washes out the information on smaller scales, which is important for the local estimation of the external convergence. Note that the difference between the H0LiCOW~III and our results for all three filtering techniques is considerably smaller than the final uncertainty on $H_0$ for the \hequad\ system alone, as quoted by \cite{2017MNRAS.465.4914B}.

\section{Testing the presence of a massive halo in the vicinity of the lensing galaxy}\label{sec:sim}

\begin{figure}
	\centering
	\captionsetup[subfigure]{labelformat=empty}
	\subfloat{
		\includegraphics[width=0.48\textwidth]{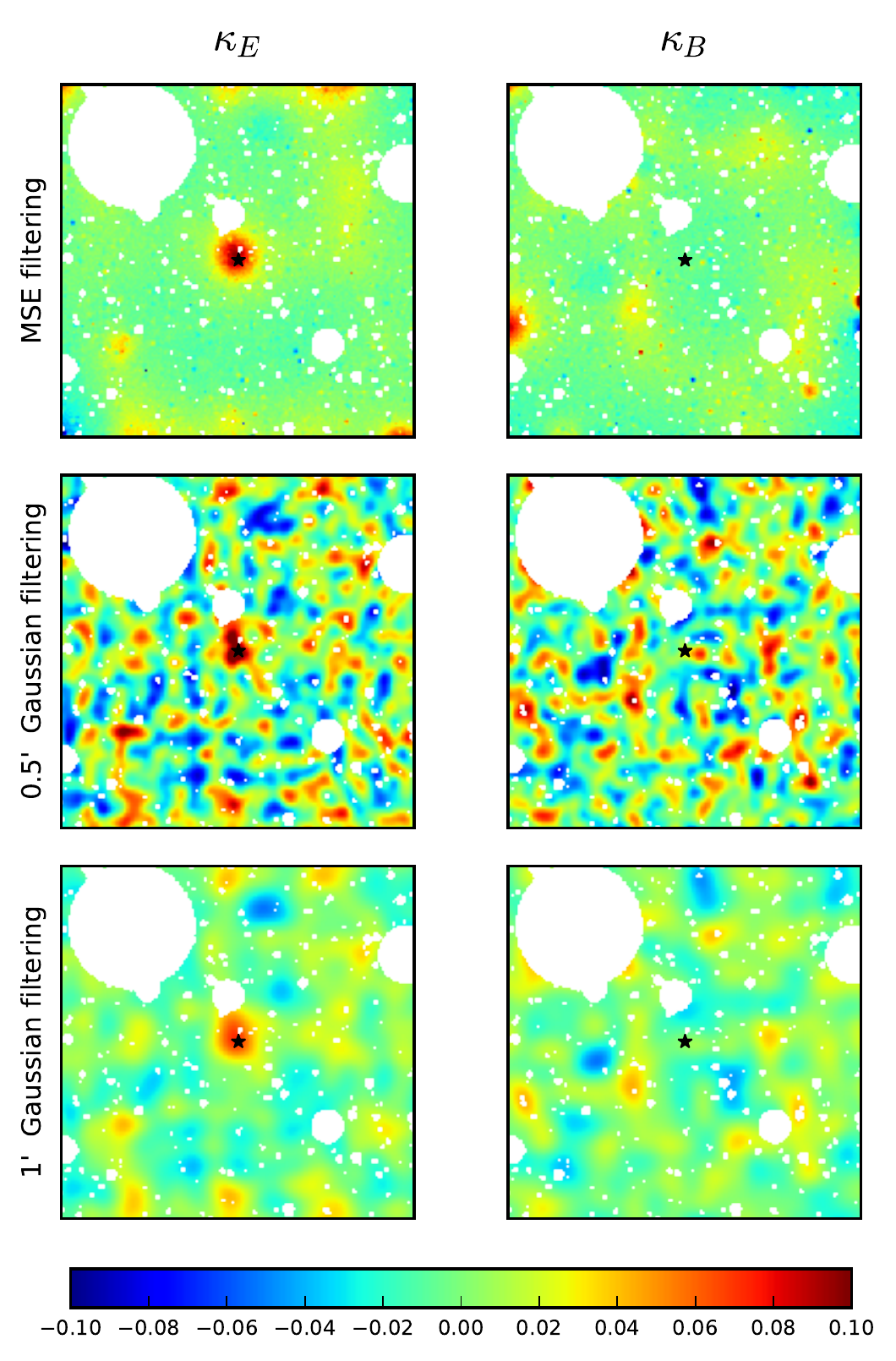}}
	\caption{Convergence maps of the simulated $0.5^{\circ} \times 0.5^{\circ}$ field containing a NFW halo with Virial mass $\mathtt{log}\, M_{vir} = 14.5 h^{-1} M_{\odot}$. Such a halo is well detected in the E-mode map but, as expected, does not show up in the B-modes. All panels are the same as in Fig.~\ref{fig:kappa}.}
	\label{fig:simulated_14.5}
\end{figure}

The lensing galaxy in \hequad\ is part of a group of galaxies \citep[e.g.][]{2017MNRAS.470.4838S}, which we do not detect in our weak lensing maps. In order to test the significance of this non-detection, and to assess the sensitivity of our mass reconstruction technique to individual halos, we perform weak lensing simulations. We simulate the shear produced by a massive halo in the vicinity of the lens and build images that mimic the data using $\tt{GalSim}$\footnote{https://github.com/GalSim-developers/GalSim} \citep{2015A&C....10..121R}. We then run our pipeline to see whether we can recover the halo injected in the image.  

\subsection{Injecting a simulated halo in the data}

The recipe we use to construct the simulated data can be summarized as follows.

\begin{itemize}
	\item {\bf Field of view:} we set the FOV to $0.5\times0.5$ deg$^2$ with a pixel size of 0.2\arcsec to mimic the \hequad\ field. The geometry of all masks is preserved;
	
	\item {\bf PSF:} for simplicity we assume a Gaussian PSF with a fixed width that matches the median width of all stars in the field of view, i.e. FWHM=0.7\arcsec;
	
	\item {\bf Galaxy population:} we leave the galaxy population the same as in the real data, i.e. with the same  ellipticities, photometric and geometric properties, the same positions on the sky, and the same photometric redshifts;
	
	\item {\bf Massive halo:} not all structures around the lensed quasar influence the $H_0$ measurements equally. In general, galaxies that lie within 1\arcmin\ radius in projection along the line of sight tend to affect $H_0$ more \cite[e.g.][]{2017ApJ...836..141M}. Similarly, structures in the foreground of the lens tend to have higher impact, while other perturbers influence $H_0$ only at the sub-percent level. Given this and the specific cumulative lensing efficiency kernel for our source galaxies, we simulate the external convergence produced by a Navarro-Frenk-White (NFW) halo \citep{1997ApJ...490..493N} at redshift $z_{halo} = 0.37$. This redshift is in the foreground of the lens and corresponds to the maximum of the cumulated efficiency kernel for our selection of source galaxies. Placing the halo at the redshift of the lens would decrease the lensing efficiency, although making the halo physically related to the lens. Our simulations and tests therefore give the minimal mass necessary to see the lensing signal at redshift $z_{halo}=0.37$. A halo at any other redshift would have to be more massive in order to be detected by our pipeline.
	
	\item {\bf Shear field:} we calculate the shear values for the simulated NFW halo at the position of all background galaxies. We use $\tt{GalSim}$ to apply the shear to the corresponding sources;
	
	\item {\bf Noise:} we add Gaussian distributed noise to the simulated field, with the same mean standard deviation as the original data.
	
\end{itemize}

Assuming a mass-concentration relation, the NFW profile can be described using only two free parameters: the Virial mass $M_{vir}$ and the redshift $z_{halo}$ \citep{2003MNRAS.344..857T}. We are using the mass-concentration relation from \cite{2008MNRAS.390L..64D} in the form
\begin{equation}
	c(z, M_{vir}) = A(M_{vir}/M_{pivot})^B(1+z)^C,
\end{equation}
where in case of the relaxed halo $M_{pivot} = 10^{12} h^{-1} M_{\odot}$, $A = 6.71$, $B = -0.091$, $C = -0.44$. We adopt the definition of the the Virial mass as the total mass within a circular area in which the mean internal density is 200 times the critical density.

To assess the sensitivity of our pipeline, we simulate 10 halos within a mass range $13.8 < \mathtt{log}\, M_{vir} [h^{-1} M_{\odot}] < 15.0$. For each halo we generate 1000 fields, where we randomize the orientation of the background galaxies before applying the corresponding shear calculated by $\tt{GalSim}$. This is necessary to cancel out any shear signal present in the data and not due to the simulated halo. Drawing 1000 simulations allows us to estimate the statistical error bars. 

We apply the pipeline described in Sect.~\ref{sec:work} to every simulated field. We first measure the shear field using the $\tt{KSB}$, and then build the corresponding convergence maps using the Kaiser\&Squires inversion, the $\tt{FASTLens}$ and the $\tt{MRLens}$ multi-resolution algorithms. An example of mass reconstruction done on one of the simulated fields is given in Fig.~\ref{fig:simulated_14.5}.

\subsection{Halo detection}

\begin{figure}
	\centering
	\includegraphics[width=0.95\columnwidth]{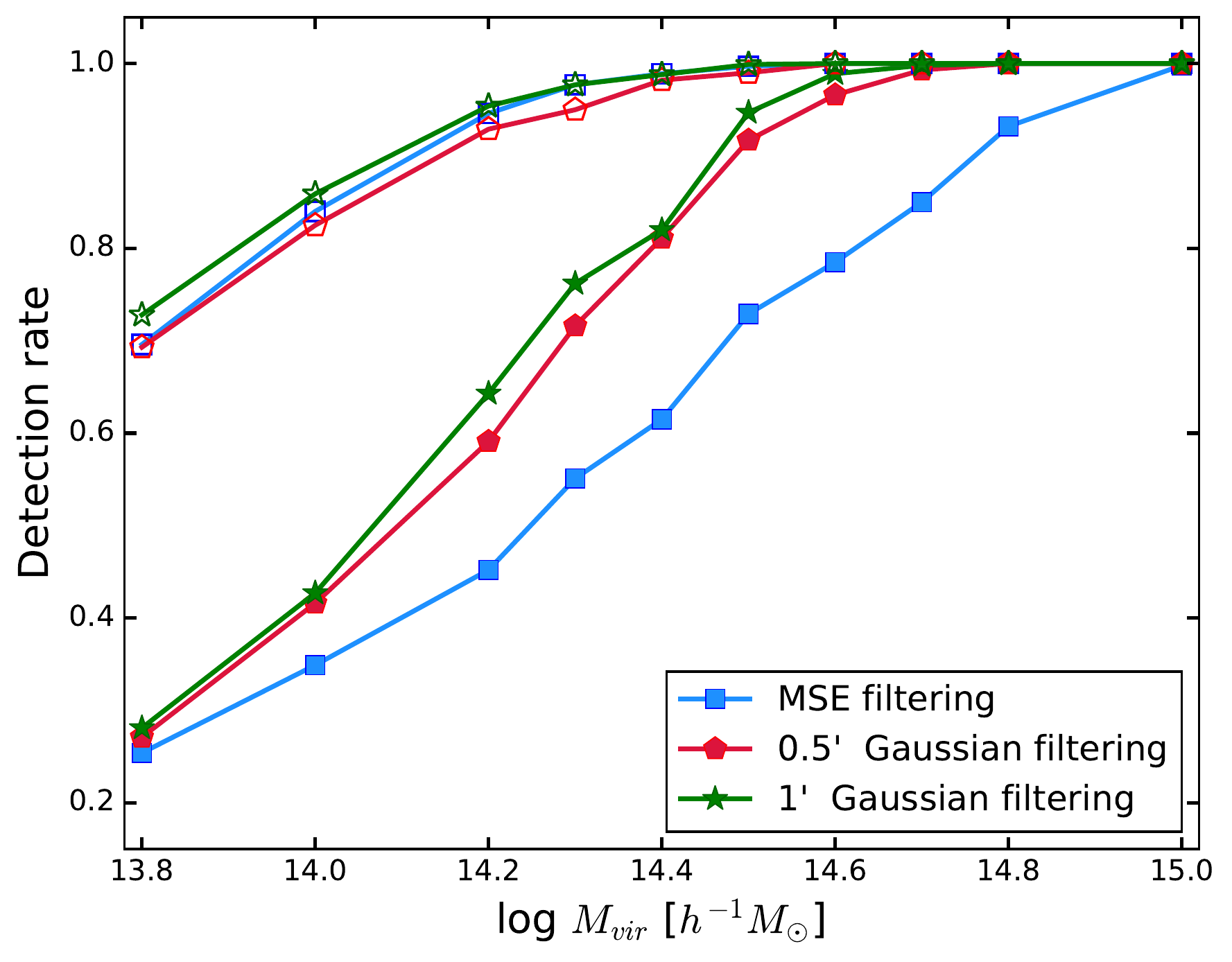}
	\caption{Detection rates of halos in the 1000 simulated fields. Filled symbols show the selection based on the S/N ratio of the halos. Open symbols show the selection based on the standard deviation of the noise in the convergence field. The color code indicates the three filtering techniques.}
	\label{fig:kappa_detection_ratio}
\end{figure}

We now test how well we can recover the structures injected in the field simulations. A natural way to detect peaks is by setting a threshold on the S/N ratio for the possible detections in each of the 1000 mocks. For the sake of comparison, we follow two different approaches. The first is based on the S/N of the peaks. We estimate the noise by taking the standard deviation of the 1000 realisations for each halo. To check if there is a halo, we calculate the signal to noise ratio of the 9 central pixels and set the threshold to $\mathtt{SNR}_9=5$. The second approach is based on the standard deviation of the noise in the fields. We sum the convergence inside the 9 central pixels, i.e. $\Sigma\kappa_9$, and compare this value to the standard deviation $\sigma_{\kappa}$ in the regions of the field that do not contain any signal from the halo. We set the criterion to $\Sigma\kappa_9>5\sigma_{\kappa}$. 

According to the central limit theorem, a Gaussian smoothing produces Gaussian noise if the number of galaxies inside the smoothing window is on average 10 or larger \citep[e.g.][]{2000MNRAS.313..524V}. With a galaxy density of 14 gals/arcmin$^2$, this criterion is satisfied for both the 0.5\arcmin\ and 1\arcmin\ Gaussian filtering. The MSE filtering, on the other hand, is a non-linear multi-scale technique, which results in the highly non-Gaussian noise \citep[e.g.][]{2011RAA....11..507J, 2016A&A...593A..88L}. Thus note that for both approaches the results for MSE filtering have to be interpreted and compared with caution, as the actual underlying statistics are non-Gaussian. However, in our case of the low S/N of the lensing signal, deviation from Gaussian statistics is not significant.

Figure~\ref{fig:kappa_detection_ratio} shows the fraction of fields with detected halos as a function of Virial mass. As expected, this fraction tends to zero for low-mass halos, and to 100\% for the high-mass end of the distribution. It is important to note that this distribution depends on the halo selection technique, with the approach based on the S/N of the peaks giving more pessimistic results. Following this approach, we are able to identify a NFW halo of $M_{vir} = 1.6 \times10^{14} h^{-1}M_{\odot}$ with a detection probability in the range $45-65\%$, depending on noise filtering technique. We consider this halo mass our detection limit at the redshift $z_{halo}=0.37$, which corresponds to the maximum of the cumulated lensing efficiency kernel, as illustrated in Fig.\ref{fig:kernel}.

For known source redshift distribution, weak lensing convergence depends on the redshift of the halo. Using the cumulated lensing efficiency kernel for our source galaxy population (see Fig.~\ref{fig:kernel} lower panel), we can rescale the convergence to match the weak lensing detection limit for other redshifts:
\begin{equation}
\kappa(z_l) = \frac{\mathcal{G}(z_l)}{\mathcal{G}(z_l=0.37)} \kappa(z_l=0.37).
\label{eq:mass_limit}
\end{equation}
The new convergence $\kappa(z_l)$ then corresponds to the Virial mass of a halo at redshift $z_l$. We calculate the limiting masses for the halos at the redshifts, where \cite{2017MNRAS.470.4838S} find spectroscopic groups in the field of \hequad. Following Table~\ref{tab:Sluse_groups}, the Virial masses of all these groups are marginally below our detection limit. \cite{2016ApJ...833..194W, 2017arXiv171009900W} estimate the Virial mass of the group at redshift $z \approx 0.18$ to be higher than that of \cite{2017MNRAS.470.4838S}, but as its mass given the error bars equals our detection limit, we can not discriminate between the two results. We can still confirm, however, that there are no structures in the field of \hequad\ more massive than our detection limits, in agreement with \cite{2017MNRAS.470.4838S}.

\begin{table}
	\begin{center}
		\begin{tabular}{ c c c | c}
			\hline
			$\bar{z}_{group}$ & $log(M_{vir}/M_{\odot})$ & $R_{vir} (\mathrm{Mpc})$ 
			& $log(M_{vir}/M_{\odot})_{limit}$ \\
			\hline
			0.0503 & 13.32 $\pm$ 0.61 & 0.635 &  15.52  \\
			0.1744 & 13.81 $\pm$ 0.40 & 1.071 &  14.57  \\
			0.1841 & 13.65 $\pm$ 0.46 & 0.954 & 14.57   \\
			0.3202 & 13.83 $\pm$ 0.36 & 1.259 & 14.37   \\
			0.4185 & 13.18 $\pm$ 0.48 & 0.873 & 14.37   \\
			0.4547 & 13.72 $\pm$ 0.36 & 1.385 & 14.42   \\
			0.5059 & 13.72 $\pm$ 0.36 & 1.373 & 14.42   \\
			0.5650 & 13.33 $\pm$ 0.43 & 0.971 & 14.52   \\
			0.7019 & 12.49 $\pm$ 0.63 & 0.654 & 14.72   \\
			\hline
		\end{tabular}
	\end{center}
	\caption{Virial mass, associated uncertainty and radius of the spectroscopic groups in the field of \hequad\ identified in \protect\cite{2017MNRAS.470.4838S}. The last column shows our weak lensing detection limit in terms of Virial mass for halos at each of the group redshifts. The limit is obtained according to Eq.~\ref{eq:mass_limit}.}
	\label{tab:Sluse_groups}
\end{table}

\section{Summary}\label{sec:summary}

We characterise the effects of the environment and the line of sight in the field of the gravitationally lensed quasar \hequad. We give a weak lensing estimation of the external convergence $\kappa_{ext}$, using the deep Subaru Suprime-Cam images and the photometric redshift catalogs from H0LiCOW~III. The weak lensing measurements are carried out on a final catalog containing 14 galaxies per square arcminute. The resulting shear field is processed with the inpainting technique to optimally account for border effects and masks of the regions affected by bright stars or extended foreground objects. 

After applying the inversion, we filter the resulting noisy convergence map in three different ways: using multi-scale entropy filtering and smoothing with 0.5\arcmin\ and 1\arcmin\ Gaussian kernels. The statistical errors are estimated using realistic mocks of the data and the systematic errors are checked by decomposing the mass maps into E- and B-modes, showing that our statistical errors dominate the systematics.

Our main result is the PDF for the external convergence inside the central pixel, i.e. at the position of the \hequad\ (see  Fig.~\ref{fig:final_result} and Table~\ref{tab:kappa_values}). We find that the \hequad\ env\&los is marginally under-dense (compared to the rest of the Universe) with a convergence being slightly negative. Our estimates are also compatible with the zero external convergence, as is found in H0LiCOW~III based on weighted galaxy number counts. Since the weak gravitational lensing measurements do not depend on assumptions about the correlation between light and mass distributions along the line of sight, our result is independent of and complementary to H0LiCOW~III.

We also test the possibility that the lensing galaxy of \hequad\ is perturbed by a single massive halo close to the lens. From the image simulations with fake halos, we show that our detection limit in mass is $M_{vir} = 1.6 \times10^{14} h^{-1}M_{\odot}$ at the redshift $z_{halo}=0.37$, which corresponds to the maximum of the cumulated lensing efficiency kernel. We scale this result according to the cumulated lensing efficiency kernel of our source galaxies in order to deduce the limit for other halo redshifts. Since we do not detect any halo in the real data, this supports the result of \cite{2017MNRAS.470.4838S}, who estimates that the Virial masses of all spectroscopic groups in the field of \hequad\ are lower than $M_{vir} = 1.6 \times10^{14} h^{-1}M_{\odot}$.

To summarise, our work supports the finding by H0LiCOW~III that the cosmology results for the \hequad\ system alone in \citet{2017MNRAS.465.4914B} are not significantly affected by line of sight effects.

\section*{Acknowledgments}
This work is supported by the Swiss National Science Foundation (SNSF). O.T. acknowledges support from the Swiss Society for Astronomy and Astrophysics (SSAA). T.T. acknowledges support by the Packard Foundation through a Packard Research Fellowship and by the National Science Foundation through grant NSF-AST1450141. S.H. acknowledges support by the DFG cluster of excellence \lq{}Origin and Structure of the Universe\rq{} (\href{http://www.universe-cluster.de}{\texttt{www.universe-cluster.de}}). S.H.S. thanks the Max Planck Society for support through the Max Planck Research Group. K.C.W. is supported by an EACOA Fellowship awarded by the East Asia Core Observatories Association, which consists of the Academia Sinica Institute of Astronomy and Astrophysics, the National Astronomical Observatory of Japan, the National Astronomical Observatories of the Chinese Academy of Sciences, and the Korea Astronomy and Space Science Institute. C.E.R. and C.D.F. acknowledge support from the National Science Foundation grant AST-1312329 and from the UC Davis Physics Department and Dean of Math and Physical Sciences. The work of P.J.M. is supported by the U.S. Department of Energy under contract number DE-AC02-76SF00515.


\bibliography{he0435wl}

\begin{thebibliography}{}
\makeatletter
\relax
\def\mn@urlcharsother{\let\do\@makeother \do\$\do\&\do\#\do\^\do\_\do\%\do\~}
\def\mn@doi{\begingroup\mn@urlcharsother \@ifnextchar [ {\mn@doi@}
  {\mn@doi@[]}}
\def\mn@doi@[#1]#2{\def\@tempa{#1}\ifx\@tempa\@empty \href
  {http://dx.doi.org/#2} {doi:#2}\else \href {http://dx.doi.org/#2} {#1}\fi
  \endgroup}
\def\mn@eprint#1#2{\mn@eprint@#1:#2::\@nil}
\def\mn@eprint@arXiv#1{\href {http://arxiv.org/abs/#1} {{\tt arXiv:#1}}}
\def\mn@eprint@dblp#1{\href {http://dblp.uni-trier.de/rec/bibtex/#1.xml}
  {dblp:#1}}
\def\mn@eprint@#1:#2:#3:#4\@nil{\def\@tempa {#1}\def\@tempb {#2}\def\@tempc
  {#3}\ifx \@tempc \@empty \let \@tempc \@tempb \let \@tempb \@tempa \fi \ifx
  \@tempb \@empty \def\@tempb {arXiv}\fi \@ifundefined
  {mn@eprint@\@tempb}{\@tempb:\@tempc}{\expandafter \expandafter \csname
  mn@eprint@\@tempb\endcsname \expandafter{\@tempc}}}

\bibitem[\protect\citeauthoryear{{Abbott} et~al.,}{{Abbott}
  et~al.}{2017}]{2017Natur.551...85A}
{Abbott} B.~P.,  et~al., 2017, \mn@doi [\nat] {10.1038/nature24471}, \href
  {http://adsabs.harvard.edu/abs/2017Natur.551...85A} {551, 85}

\bibitem[\protect\citeauthoryear{{Alam} et~al.,}{{Alam}
  et~al.}{2017}]{2017MNRAS.470.2617A}
{Alam} S.,  et~al., 2017, \mn@doi [\mnras] {10.1093/mnras/stx721}, \href
  {http://adsabs.harvard.edu/abs/2017MNRAS.470.2617A} {470, 2617}

\bibitem[\protect\citeauthoryear{{Auger}, {Fassnacht}, {Abrahamse}, {Lubin}  \&
  {Squires}}{{Auger} et~al.}{2007}]{2007AJ....134..668A}
{Auger} M.~W.,  {Fassnacht} C.~D.,  {Abrahamse} A.~L.,  {Lubin} L.~M.,
  {Squires} G.~K.,  2007, \mn@doi [\aj] {10.1086/519238}, \href
  {http://adsabs.harvard.edu/abs/2007AJ....134..668A} {134, 668}

\bibitem[\protect\citeauthoryear{{Bartelmann} \& {Schneider}}{{Bartelmann} \&
  {Schneider}}{2001}]{2001PhR...340..291B}
{Bartelmann} M.,  {Schneider} P.,  2001, \mn@doi [\physrep]
  {10.1016/S0370-1573(00)00082-X}, \href
  {http://adsabs.harvard.edu/abs/2001PhR...340..291B} {340, 291}

\bibitem[\protect\citeauthoryear{{Benjamini} \& {Hochberg}}{{Benjamini} \&
  {Hochberg}}{1995}]{fdr}
{Benjamini} Y.,  {Hochberg} Y.,  1995, \mn@doi [J. R. Stat. Soc. B]
  {1510.03103}, \href {http://adsabs.harvard.edu/abs/2015arXiv151003103G} {57,
  289}

\bibitem[\protect\citeauthoryear{{Bertin} \& {Arnouts}}{{Bertin} \&
  {Arnouts}}{1996}]{1996A&AS..117..393B}
{Bertin} E.,  {Arnouts} S.,  1996, \mn@doi [\aaps] {10.1051/aas:1996164}, \href
  {http://adsabs.harvard.edu/abs/1996A%26AS..117..393B} {117, 393}

\bibitem[\protect\citeauthoryear{{Birrer}, {Welschen}, {Amara}  \&
  {Refregier}}{{Birrer} et~al.}{2017}]{2017JCAP...04..049B}
{Birrer} S.,  {Welschen} C.,  {Amara} A.,   {Refregier} A.,  2017, \mn@doi
  [\jcap] {10.1088/1475-7516/2017/04/049}, \href
  {http://adsabs.harvard.edu/abs/2017JCAP...04..049B} {4, 049}

\bibitem[\protect\citeauthoryear{{Bonvin}, {Tewes}, {Courbin}, {Kuntzer},
  {Sluse}  \& {Meylan}}{{Bonvin} et~al.}{2016}]{2016A&A...585A..88B}
{Bonvin} V.,  {Tewes} M.,  {Courbin} F.,  {Kuntzer} T.,  {Sluse} D.,   {Meylan}
  G.,  2016, \mn@doi [\aap] {10.1051/0004-6361/201526704}, \href
  {http://adsabs.harvard.edu/abs/2016A%26A...585A..88B} {585, A88}

\bibitem[\protect\citeauthoryear{{Bonvin} et~al.,}{{Bonvin}
  et~al.}{2017}]{2017MNRAS.465.4914B}
{Bonvin} V.,  et~al., 2017, \mn@doi [\mnras] {10.1093/mnras/stw3006}, \href
  {http://adsabs.harvard.edu/abs/2017MNRAS.465.4914B} {465, 4914}

\bibitem[\protect\citeauthoryear{{Coe} \& {Moustakas}}{{Coe} \&
  {Moustakas}}{2009}]{2009ApJ...706...45C}
{Coe} D.,  {Moustakas} L.~A.,  2009, \mn@doi [\apj]
  {10.1088/0004-637X/706/1/45}, \href
  {http://adsabs.harvard.edu/abs/2009ApJ...706...45C} {706, 45}

\bibitem[\protect\citeauthoryear{{Collett} \& {Cunnington}}{{Collett} \&
  {Cunnington}}{2016}]{2016arXiv160508341C}
{Collett} T.~E.,  {Cunnington} S.~D.,  2016, \mn@doi [\mnras]
  {10.1093/mnras/stw1856}, \href
  {http://adsabs.harvard.edu/abs/2016MNRAS.462.3255C} {462, 3255}

\bibitem[\protect\citeauthoryear{{Collett} et~al.,}{{Collett}
  et~al.}{2013}]{2013MNRAS.432..679C}
{Collett} T.~E.,  et~al., 2013, \mn@doi [\mnras] {10.1093/mnras/stt504}, \href
  {http://adsabs.harvard.edu/abs/2013MNRAS.432..679C} {432, 679}

\bibitem[\protect\citeauthoryear{{Courbin}, {Eigenbrod}, {Vuissoz}, {Meylan}
  \& {Magain}}{{Courbin} et~al.}{2005}]{2005IAUS..225..297C}
{Courbin} F.,  {Eigenbrod} A.,  {Vuissoz} C.,  {Meylan} G.,   {Magain} P.,
  2005, in {Mellier} Y.,  {Meylan} G.,  eds,  IAU Symposium Vol. 225,
  Gravitational Lensing Impact on Cosmology. pp 297--303,
  \mn@doi{10.1017/S1743921305002097}

\bibitem[\protect\citeauthoryear{{Courbin} et~al.,}{{Courbin}
  et~al.}{2011}]{2011A&A...536A..53C}
{Courbin} F.,  et~al., 2011, \mn@doi [\aap] {10.1051/0004-6361/201015709},
  \href {http://adsabs.harvard.edu/abs/2011A%26A...536A..53C} {536, A53}

\bibitem[\protect\citeauthoryear{{DES Collaboration} et~al.,}{{DES
  Collaboration} et~al.}{2017}]{2017arXiv171100403D}
{DES Collaboration} et~al., 2017, preprint, \href
  {http://adsabs.harvard.edu/abs/2017arXiv171100403D} {} (\mn@eprint {arXiv}
  {1711.00403})

\bibitem[\protect\citeauthoryear{{De Lucia} \& {Blaizot}}{{De Lucia} \&
  {Blaizot}}{2007}]{2007MNRAS.375....2D}
{De Lucia} G.,  {Blaizot} J.,  2007, \mn@doi [\mnras]
  {10.1111/j.1365-2966.2006.11287.x}, \href
  {http://adsabs.harvard.edu/abs/2007MNRAS.375....2D} {375, 2}

\bibitem[\protect\citeauthoryear{{Ding} et~al.,}{{Ding}
  et~al.}{2017a}]{2017MNRAS.465.4634D}
{Ding} X.,  et~al., 2017a, \mn@doi [\mnras] {10.1093/mnras/stw3078}, \href
  {http://adsabs.harvard.edu/abs/2017MNRAS.465.4634D} {465, 4634}

\bibitem[\protect\citeauthoryear{{Ding} et~al.,}{{Ding}
  et~al.}{2017b}]{2017MNRAS.472...90D}
{Ding} X.,  et~al., 2017b, \mn@doi [\mnras] {10.1093/mnras/stx1972}, \href
  {http://adsabs.harvard.edu/abs/2017MNRAS.472...90D} {472, 90}

\bibitem[\protect\citeauthoryear{{Duffy}, {Schaye}, {Kay}  \& {Dalla
  Vecchia}}{{Duffy} et~al.}{2008}]{2008MNRAS.390L..64D}
{Duffy} A.~R.,  {Schaye} J.,  {Kay} S.~T.,   {Dalla Vecchia} C.,  2008, \mn@doi
  [\mnras] {10.1111/j.1745-3933.2008.00537.x}, \href
  {http://adsabs.harvard.edu/abs/2008MNRAS.390L..64D} {390, L64}

\bibitem[\protect\citeauthoryear{{Eigenbrod}, {Courbin}, {Meylan}, {Vuissoz}
  \& {Magain}}{{Eigenbrod} et~al.}{2006}]{2006A&A...451..759E}
{Eigenbrod} A.,  {Courbin} F.,  {Meylan} G.,  {Vuissoz} C.,   {Magain} P.,
  2006, \mn@doi [\aap] {10.1051/0004-6361:20054454}, \href
  {http://adsabs.harvard.edu/abs/2006A%26A...451..759E} {451, 759}

\bibitem[\protect\citeauthoryear{{Elad}, {Starck}, {Querre}  \&
  {Donoho}}{{Elad} et~al.}{2005}]{elad2005}
{Elad} M.,  {Starck} J.-L.,  {Querre} P.,   {Donoho} D.,  2005, \mn@doi [J. on
  Applied and Computational Harmonic Analysis] {10.1016/j.acha.2005.03.005},
  \href
  {http://ac.els-cdn.com/S1063520305000655/1-s2.0-S1063520305000655-main.pdf?_tid=83981e26-79c0-11e6-9051-00000aab0f01&acdnat=1473778023_f3209bc9051b569407b08f61401f90d7}
  {19, 340}

\bibitem[\protect\citeauthoryear{{Fadely}, {Keeton}, {Nakajima}  \&
  {Bernstein}}{{Fadely} et~al.}{2010}]{2010ApJ...711..246F}
{Fadely} R.,  {Keeton} C.~R.,  {Nakajima} R.,   {Bernstein} G.~M.,  2010,
  \mn@doi [\apj] {10.1088/0004-637X/711/1/246}, \href
  {http://adsabs.harvard.edu/abs/2010ApJ...711..246F} {711, 246}

\bibitem[\protect\citeauthoryear{{Falco}, {Gorenstein}  \& {Shapiro}}{{Falco}
  et~al.}{1985}]{1985ApJ...289L...1F}
{Falco} E.~E.,  {Gorenstein} M.~V.,   {Shapiro} I.~I.,  1985, \mn@doi [\apjl]
  {10.1086/184422}, \href {http://adsabs.harvard.edu/abs/1985ApJ...289L...1F}
  {289, L1}

\bibitem[\protect\citeauthoryear{{Fassnacht}, {Xanthopoulos}, {Koopmans}  \&
  {Rusin}}{{Fassnacht} et~al.}{2002}]{2002ApJ...581..823F}
{Fassnacht} C.~D.,  {Xanthopoulos} E.,  {Koopmans} L.~V.~E.,   {Rusin} D.,
  2002, \mn@doi [\apj] {10.1086/344368}, \href
  {http://adsabs.harvard.edu/abs/2002ApJ...581..823F} {581, 823}

\bibitem[\protect\citeauthoryear{{Fassnacht}, {Gal}, {Lubin}, {McKean},
  {Squires}  \& {Readhead}}{{Fassnacht} et~al.}{2006}]{2006ApJ...642...30F}
{Fassnacht} C.~D.,  {Gal} R.~R.,  {Lubin} L.~M.,  {McKean} J.~P.,  {Squires}
  G.~K.,   {Readhead} A.~C.~S.,  2006, \mn@doi [\apj] {10.1086/500927}, \href
  {http://adsabs.harvard.edu/abs/2006ApJ...642...30F} {642, 30}

\bibitem[\protect\citeauthoryear{{Fassnacht}, {Koopmans}  \&
  {Wong}}{{Fassnacht} et~al.}{2011}]{2011MNRAS.410.2167F}
{Fassnacht} C.~D.,  {Koopmans} L.~V.~E.,   {Wong} K.~C.,  2011, \mn@doi
  [\mnras] {10.1111/j.1365-2966.2010.17591.x}, \href
  {http://adsabs.harvard.edu/abs/2011MNRAS.410.2167F} {410, 2167}

\bibitem[\protect\citeauthoryear{{Fern{\'a}ndez-Arenas}
  et~al.,}{{Fern{\'a}ndez-Arenas} et~al.}{2017}]{2017arXiv171005951F}
{Fern{\'a}ndez-Arenas} D.,  et~al., 2017, preprint, \href
  {http://adsabs.harvard.edu/abs/2017arXiv171005951F} {} (\mn@eprint {arXiv}
  {1710.05951})

\bibitem[\protect\citeauthoryear{{Fischer}, {Bernstein}, {Rhee}  \&
  {Tyson}}{{Fischer} et~al.}{1997}]{1997AJ....113..521F}
{Fischer} P.,  {Bernstein} G.,  {Rhee} G.,   {Tyson} J.~A.,  1997, \mn@doi
  [\aj] {10.1086/118272}, \href
  {http://adsabs.harvard.edu/abs/1997AJ....113..521F} {113, 521}

\bibitem[\protect\citeauthoryear{{Freedman}, {Madore}, {Scowcroft}, {Burns},
  {Monson}, {Persson}, {Seibert}  \& {Rigby}}{{Freedman}
  et~al.}{2012}]{2012ApJ...758...24F}
{Freedman} W.~L.,  {Madore} B.~F.,  {Scowcroft} V.,  {Burns} C.,  {Monson} A.,
  {Persson} S.~E.,  {Seibert} M.,   {Rigby} J.,  2012, \mn@doi [\apj]
  {10.1088/0004-637X/758/1/24}, \href
  {http://adsabs.harvard.edu/abs/2012ApJ...758...24F} {758, 24}

\bibitem[\protect\citeauthoryear{{Garc{\'{\i}}a} et~al.,}{{Garc{\'{\i}}a}
  et~al.}{2014}]{2014A&A...568A..10G}
{Garc{\'{\i}}a} R.~A.,  et~al., 2014, \mn@doi [\aap]
  {10.1051/0004-6361/201323326}, \href
  {http://adsabs.harvard.edu/abs/2014A%26A...568A..10G} {568, A10}

\bibitem[\protect\citeauthoryear{{Goobar} et~al.,}{{Goobar}
  et~al.}{2017}]{2016arXiv161100014G}
{Goobar} A.,  et~al., 2017, \mn@doi [Science] {10.1126/science.aal2729}, \href
  {http://adsabs.harvard.edu/abs/2017Sci...356..291G} {356, 291}

\bibitem[\protect\citeauthoryear{{Greene} et~al.,}{{Greene}
  et~al.}{2013}]{2013ApJ...768...39G}
{Greene} Z.~S.,  et~al., 2013, \mn@doi [\apj] {10.1088/0004-637X/768/1/39},
  \href {http://adsabs.harvard.edu/abs/2013ApJ...768...39G} {768, 39}

\bibitem[\protect\citeauthoryear{{Heymans} et~al.,}{{Heymans}
  et~al.}{2006}]{2006MNRAS.368.1323H}
{Heymans} C.,  et~al., 2006, \mn@doi [\mnras]
  {10.1111/j.1365-2966.2006.10198.x}, \href
  {http://adsabs.harvard.edu/abs/2006MNRAS.368.1323H} {368, 1323}

\bibitem[\protect\citeauthoryear{{Hilbert}, {White}, {Hartlap}  \&
  {Schneider}}{{Hilbert} et~al.}{2008}]{2008MNRAS.386.1845H}
{Hilbert} S.,  {White} S.~D.~M.,  {Hartlap} J.,   {Schneider} P.,  2008,
  \mn@doi [\mnras] {10.1111/j.1365-2966.2008.13190.x}, \href
  {http://adsabs.harvard.edu/abs/2008MNRAS.386.1845H} {386, 1845}

\bibitem[\protect\citeauthoryear{{Hilbert}, {Hartlap}, {White}  \&
  {Schneider}}{{Hilbert} et~al.}{2009a}]{2009A&A...499...31H}
{Hilbert} S.,  {Hartlap} J.,  {White} S.~D.~M.,   {Schneider} P.,  2009a,
  \mn@doi [\aap] {10.1051/0004-6361/200811054}, \href
  {http://adsabs.harvard.edu/abs/2009A%26A...499...31H} {499, 31}

\bibitem[\protect\citeauthoryear{{Hilbert}, {Hartlap}, {White}  \&
  {Schneider}}{{Hilbert} et~al.}{2009b}]{2009AA...499...31H}
{Hilbert} S.,  {Hartlap} J.,  {White} S.~D.~M.,   {Schneider} P.,  2009b,
  \mn@doi [\aap] {10.1051/0004-6361/200811054}, \href
  {http://adsabs.harvard.edu/abs/2009A%26A...499...31H} {499, 31}

\bibitem[\protect\citeauthoryear{{Holschneider}, {Kronland-Martinet}, {Morlet}
  \& {Tchamitchian}}{{Holschneider} et~al.}{1989}]{1989wtfm.conf..286H}
{Holschneider} M.,  {Kronland-Martinet} R.,  {Morlet} J.,   {Tchamitchian} P.,
  1989, in {Combes} J.-M.,  {Grossmann} A.,   {Tchamitchian} P.,  eds,
  Wavelets. Time-Frequency Methods and Phase Space. p.~286

\bibitem[\protect\citeauthoryear{{Jaroszy{\'n}ski} \&
  {Skowron}}{{Jaroszy{\'n}ski} \& {Skowron}}{2016}]{2016MNRAS.462.1405J}
{Jaroszy{\'n}ski} M.,  {Skowron} J.,  2016, \mn@doi [\mnras]
  {10.1093/mnras/stw1739}, \href
  {http://adsabs.harvard.edu/abs/2016MNRAS.462.1405J} {462, 1405}

\bibitem[\protect\citeauthoryear{{Jiao}, {Shan}  \& {Fan}}{{Jiao}
  et~al.}{2011}]{2011RAA....11..507J}
{Jiao} Y.-X.,  {Shan} H.-Y.,   {Fan} Z.-H.,  2011, \mn@doi [Research in
  Astronomy and Astrophysics] {10.1088/1674-4527/11/5/002}, \href
  {http://adsabs.harvard.edu/abs/2011RAA....11..507J} {11, 507}

\bibitem[\protect\citeauthoryear{{Kaiser} \& {Squires}}{{Kaiser} \&
  {Squires}}{1993}]{1993ApJ...404..441K}
{Kaiser} N.,  {Squires} G.,  1993, \mn@doi [\apj] {10.1086/172297}, \href
  {http://adsabs.harvard.edu/abs/1993ApJ...404..441K} {404, 441}

\bibitem[\protect\citeauthoryear{{Kaiser}, {Squires}  \& {Broadhurst}}{{Kaiser}
  et~al.}{1995}]{1995ApJ...449..460K}
{Kaiser} N.,  {Squires} G.,   {Broadhurst} T.,  1995, \mn@doi [\apj]
  {10.1086/176071}, \href {http://adsabs.harvard.edu/abs/1995ApJ...449..460K}
  {449, 460}

\bibitem[\protect\citeauthoryear{{Keeton} \& {Zabludoff}}{{Keeton} \&
  {Zabludoff}}{2004}]{2004ApJ...612..660K}
{Keeton} C.~R.,  {Zabludoff} A.~I.,  2004, \mn@doi [\apj] {10.1086/422745},
  \href {http://adsabs.harvard.edu/abs/2004ApJ...612..660K} {612, 660}

\bibitem[\protect\citeauthoryear{{Kelly} et~al.,}{{Kelly}
  et~al.}{2015}]{2015Sci...347.1123K}
{Kelly} P.~L.,  et~al., 2015, \mn@doi [Science] {10.1126/science.aaa3350},
  \href {http://adsabs.harvard.edu/abs/2015Sci...347.1123K} {347, 1123}

\bibitem[\protect\citeauthoryear{{Koopmans}}{{Koopmans}}{2004}]{2004astro.ph.12596K}
{Koopmans} L.~V.~E.,  2004, ArXiv Astrophysics e-prints, \href
  {http://adsabs.harvard.edu/abs/2004astro.ph.12596K} {}

\bibitem[\protect\citeauthoryear{{Lehar}, {Hewitt}, {Burke}  \&
  {Roberts}}{{Lehar} et~al.}{1992}]{1992ApJ...384..453L}
{Lehar} J.,  {Hewitt} J.~N.,  {Burke} B.~F.,   {Roberts} D.~H.,  1992, \mn@doi
  [\apj] {10.1086/170887}, \href
  {http://adsabs.harvard.edu/abs/1992ApJ...384..453L} {384, 453}

\bibitem[\protect\citeauthoryear{{Lin}, {Kilbinger}  \& {Pires}}{{Lin}
  et~al.}{2016}]{2016A&A...593A..88L}
{Lin} C.-A.,  {Kilbinger} M.,   {Pires} S.,  2016, \mn@doi [\aap]
  {10.1051/0004-6361/201628565}, \href
  {http://adsabs.harvard.edu/abs/2016A%26A...593A..88L} {593, A88}

\bibitem[\protect\citeauthoryear{{Marshall}, {Rajguru}  \& {Slosar}}{{Marshall}
  et~al.}{2006}]{2006PhRvD..73f7302M}
{Marshall} P.,  {Rajguru} N.,   {Slosar} A.,  2006, \mn@doi [\prd]
  {10.1103/PhysRevD.73.067302}, \href
  {http://adsabs.harvard.edu/abs/2006PhRvD..73f7302M} {73, 067302}

\bibitem[\protect\citeauthoryear{{McCully}, {Keeton}, {Wong}  \&
  {Zabludoff}}{{McCully} et~al.}{2014}]{2014MNRAS.443.3631M}
{McCully} C.,  {Keeton} C.~R.,  {Wong} K.~C.,   {Zabludoff} A.~I.,  2014,
  \mn@doi [\mnras] {10.1093/mnras/stu1316}, \href
  {http://adsabs.harvard.edu/abs/2014MNRAS.443.3631M} {443, 3631}

\bibitem[\protect\citeauthoryear{{McCully}, {Keeton}, {Wong}  \&
  {Zabludoff}}{{McCully} et~al.}{2017a}]{2016arXiv160105417M}
{McCully} C.,  {Keeton} C.~R.,  {Wong} K.~C.,   {Zabludoff} A.~I.,  2017a,
  \mn@doi [\apj] {10.3847/1538-4357/836/1/141}, \href
  {http://adsabs.harvard.edu/abs/2017ApJ...836..141M} {836, 141}

\bibitem[\protect\citeauthoryear{{McCully}, {Keeton}, {Wong}  \&
  {Zabludoff}}{{McCully} et~al.}{2017b}]{2017ApJ...836..141M}
{McCully} C.,  {Keeton} C.~R.,  {Wong} K.~C.,   {Zabludoff} A.~I.,  2017b,
  \mn@doi [\apj] {10.3847/1538-4357/836/1/141}, \href
  {http://adsabs.harvard.edu/abs/2017ApJ...836..141M} {836, 141}

\bibitem[\protect\citeauthoryear{{Miller} et~al.,}{{Miller}
  et~al.}{2001}]{2001AJ....122.3492M}
{Miller} C.~J.,  et~al., 2001, \mn@doi [\aj] {10.1086/324109}, \href
  {http://adsabs.harvard.edu/abs/2001AJ....122.3492M} {122, 3492}

\bibitem[\protect\citeauthoryear{{Momcheva}, {Williams}, {Keeton}  \&
  {Zabludoff}}{{Momcheva} et~al.}{2006}]{2006ApJ...641..169M}
{Momcheva} I.,  {Williams} K.,  {Keeton} C.,   {Zabludoff} A.,  2006, \mn@doi
  [\apj] {10.1086/500382}, \href
  {http://adsabs.harvard.edu/abs/2006ApJ...641..169M} {641, 169}

\bibitem[\protect\citeauthoryear{{Morgan}, {Kochanek}, {Pevunova}  \&
  {Schechter}}{{Morgan} et~al.}{2005}]{2005AJ....129.2531M}
{Morgan} N.~D.,  {Kochanek} C.~S.,  {Pevunova} O.,   {Schechter} P.~L.,  2005,
  \mn@doi [\aj] {10.1086/430145}, \href
  {http://adsabs.harvard.edu/abs/2005AJ....129.2531M} {129, 2531}

\bibitem[\protect\citeauthoryear{{Nakajima}, {Bernstein}, {Fadely}, {Keeton}
  \& {Schrabback}}{{Nakajima} et~al.}{2009}]{2009ApJ...697.1793N}
{Nakajima} R.,  {Bernstein} G.~M.,  {Fadely} R.,  {Keeton} C.~R.,
  {Schrabback} T.,  2009, \mn@doi [\apj] {10.1088/0004-637X/697/2/1793}, \href
  {http://adsabs.harvard.edu/abs/2009ApJ...697.1793N} {697, 1793}

\bibitem[\protect\citeauthoryear{{Navarro}, {Frenk}  \& {White}}{{Navarro}
  et~al.}{1997}]{1997ApJ...490..493N}
{Navarro} J.~F.,  {Frenk} C.~S.,   {White} S.~D.~M.,  1997, \apj, \href
  {http://adsabs.harvard.edu/abs/1997ApJ...490..493N} {490, 493}

\bibitem[\protect\citeauthoryear{{Perotto}, {Bobin}, {Plaszczynski}, {Starck}
  \& {Lavabre}}{{Perotto} et~al.}{2010}]{2010A&A...519A...4P}
{Perotto} L.,  {Bobin} J.,  {Plaszczynski} S.,  {Starck} J.-L.,   {Lavabre} A.,
   2010, \mn@doi [\aap] {10.1051/0004-6361/200912001}, \href
  {http://adsabs.harvard.edu/abs/2010A%26A...519A...4P} {519, A4}

\bibitem[\protect\citeauthoryear{{Pires}, {Starck}, {Amara}, {Teyssier},
  {R{\'e}fr{\'e}gier}  \& {Fadili}}{{Pires}
  et~al.}{2009a}]{2009MNRAS.395.1265P}
{Pires} S.,  {Starck} J.-L.,  {Amara} A.,  {Teyssier} R.,  {R{\'e}fr{\'e}gier}
  A.,   {Fadili} J.,  2009a, \mn@doi [\mnras]
  {10.1111/j.1365-2966.2009.14625.x}, \href
  {http://adsabs.harvard.edu/abs/2009MNRAS.395.1265P} {395, 1265}

\bibitem[\protect\citeauthoryear{{Pires}, {Starck}, {Amara},
  {R{\'e}fr{\'e}gier}  \& {Teyssier}}{{Pires}
  et~al.}{2009b}]{2009A&A...505..969P}
{Pires} S.,  {Starck} J.-L.,  {Amara} A.,  {R{\'e}fr{\'e}gier} A.,   {Teyssier}
  R.,  2009b, \mn@doi [\aap] {10.1051/0004-6361/200811459}, \href
  {http://adsabs.harvard.edu/abs/2009A%26A...505..969P} {505, 969}

\bibitem[\protect\citeauthoryear{{Pires}, {Mathur}, {Garc{\'{\i}}a}, {Ballot},
  {Stello}  \& {Sato}}{{Pires} et~al.}{2015}]{2015A&A...574A..18P}
{Pires} S.,  {Mathur} S.,  {Garc{\'{\i}}a} R.~A.,  {Ballot} J.,  {Stello} D.,
  {Sato} K.,  2015, \mn@doi [\aap] {10.1051/0004-6361/201322361}, \href
  {http://adsabs.harvard.edu/abs/2015A%26A...574A..18P} {574, A18}

\bibitem[\protect\citeauthoryear{{Planck Collaboration} et~al.,}{{Planck
  Collaboration} et~al.}{2016}]{2016A&A...594A..13P}
{Planck Collaboration} et~al., 2016, \mn@doi [\aap]
  {10.1051/0004-6361/201525830}, \href
  {http://adsabs.harvard.edu/abs/2016A%26A...594A..13P} {594, A13}

\bibitem[\protect\citeauthoryear{{Plaszczynski}, {Lavabre}, {Perotto}  \&
  {Starck}}{{Plaszczynski} et~al.}{2012}]{2012A&A...544A..27P}
{Plaszczynski} S.,  {Lavabre} A.,  {Perotto} L.,   {Starck} J.-L.,  2012,
  \mn@doi [\aap] {10.1051/0004-6361/201218899}, \href
  {http://adsabs.harvard.edu/abs/2012A%26A...544A..27P} {544, A27}

\bibitem[\protect\citeauthoryear{{Refsdal}}{{Refsdal}}{1964}]{Refsdal64}
{Refsdal} S.,  1964, \mnras, \href
  {http://adsabs.harvard.edu/cgi-bin/nph-bib_query?bibcode=1964MNRAS.128..307R&db_key=AST}
  {128, 307}

\bibitem[\protect\citeauthoryear{{Reid}, {Braatz}, {Condon}, {Lo}, {Kuo},
  {Impellizzeri}  \& {Henkel}}{{Reid} et~al.}{2013}]{2013ApJ...767..154R}
{Reid} M.~J.,  {Braatz} J.~A.,  {Condon} J.~J.,  {Lo} K.~Y.,  {Kuo} C.~Y.,
  {Impellizzeri} C.~M.~V.,   {Henkel} C.,  2013, \mn@doi [\apj]
  {10.1088/0004-637X/767/2/154}, \href
  {http://adsabs.harvard.edu/abs/2013ApJ...767..154R} {767, 154}

\bibitem[\protect\citeauthoryear{{Riess} et~al.,}{{Riess}
  et~al.}{2016}]{2016ApJ...826...56R}
{Riess} A.~G.,  et~al., 2016, \mn@doi [\apj] {10.3847/0004-637X/826/1/56},
  \href {http://adsabs.harvard.edu/abs/2016ApJ...826...56R} {826, 56}

\bibitem[\protect\citeauthoryear{{Rodney} et~al.,}{{Rodney}
  et~al.}{2015}]{2015ApJ...811...70R}
{Rodney} S.~A.,  et~al., 2015, \mn@doi [\apj] {10.1088/0004-637X/811/1/70},
  \href {http://adsabs.harvard.edu/abs/2015ApJ...811...70R} {811, 70}

\bibitem[\protect\citeauthoryear{{Rowe} et~al.,}{{Rowe}
  et~al.}{2015}]{2015A&C....10..121R}
{Rowe} B.~T.~P.,  et~al., 2015, \mn@doi [Astronomy and Computing]
  {10.1016/j.ascom.2015.02.002}, \href
  {http://adsabs.harvard.edu/abs/2015A%26C....10..121R} {10, 121}

\bibitem[\protect\citeauthoryear{{Rusu} et~al.,}{{Rusu}
  et~al.}{2017}]{2016arXiv160701047R}
{Rusu} C.~E.,  et~al., 2017, \mn@doi [\mnras] {10.1093/mnras/stx285}, \href
  {http://adsabs.harvard.edu/abs/2017MNRAS.467.4220R} {467, 4220}

\bibitem[\protect\citeauthoryear{{Schild}}{{Schild}}{1990}]{1990AJ....100.1771S}
{Schild} R.~E.,  1990, \mn@doi [\aj] {10.1086/115634}, \href
  {http://adsabs.harvard.edu/abs/1990AJ....100.1771S} {100, 1771}

\bibitem[\protect\citeauthoryear{{Schneider} \& {Sluse}}{{Schneider} \&
  {Sluse}}{2013}]{2013A&A...559A..37S}
{Schneider} P.,  {Sluse} D.,  2013, \mn@doi [\aap]
  {10.1051/0004-6361/201321882}, \href
  {http://adsabs.harvard.edu/abs/2013A%26A...559A..37S} {559, A37}

\bibitem[\protect\citeauthoryear{{Schneider}, {Kochanek}  \&
  {Wambsganss}}{{Schneider} et~al.}{2006}]{schneider2006}
{Schneider} P.,  {Kochanek} C.,   {Wambsganss} J.,  2006, Gravitational
  Lensing: Strong, Weak and Micro.
Springer-Verlag Berlin Heidelberg, DOI 10.1007/978-3-540-30310-7

\bibitem[\protect\citeauthoryear{{Sluse}, {Hutsem{\'e}kers}, {Courbin},
  {Meylan}  \& {Wambsganss}}{{Sluse} et~al.}{2012}]{2012A&A...544A..62S}
{Sluse} D.,  {Hutsem{\'e}kers} D.,  {Courbin} F.,  {Meylan} G.,   {Wambsganss}
  J.,  2012, \mn@doi [\aap] {10.1051/0004-6361/201219125}, \href
  {http://adsabs.harvard.edu/abs/2012A%26A...544A..62S} {544, A62}

\bibitem[\protect\citeauthoryear{{Sluse} et~al.,}{{Sluse}
  et~al.}{2017}]{2017MNRAS.470.4838S}
{Sluse} D.,  et~al., 2017, \mn@doi [\mnras] {10.1093/mnras/stx1484}, \href
  {http://adsabs.harvard.edu/abs/2017MNRAS.470.4838S} {470, 4838}

\bibitem[\protect\citeauthoryear{{Springel}}{{Springel}}{2005}]{2005MNRAS.364.1105S}
{Springel} V.,  2005, \mn@doi [\mnras] {10.1111/j.1365-2966.2005.09655.x},
  \href {http://adsabs.harvard.edu/abs/2005MNRAS.364.1105S} {364, 1105}

\bibitem[\protect\citeauthoryear{{Springel} et~al.,}{{Springel}
  et~al.}{2005}]{2005Natur.435..629S}
{Springel} V.,  et~al., 2005, \mn@doi [\nat] {10.1038/nature03597}, \href
  {http://adsabs.harvard.edu/abs/2005Natur.435..629S} {435, 629}

\bibitem[\protect\citeauthoryear{{Starck} \& {Murtagh}}{{Starck} \&
  {Murtagh}}{2006}]{2006aida.book.....S}
{Starck} J.-L.,  {Murtagh} F.,  2006, {Astronomical Image and Data Analysis},
  \mn@doi{10.1007/978-3-540-33025-7.
}

\bibitem[\protect\citeauthoryear{{Starck}, {Murtagh}, {Pirenne}  \&
  {Albrecht}}{{Starck} et~al.}{1996}]{1996PASP..108..446S}
{Starck} J.-L.,  {Murtagh} F.,  {Pirenne} B.,   {Albrecht} M.,  1996, \mn@doi
  [\pasp] {10.1086/133746}, \href
  {http://adsabs.harvard.edu/abs/1996PASP..108..446S} {108, 446}

\bibitem[\protect\citeauthoryear{{Starck}, {Murtagh}, {Querre}  \&
  {Bonnarel}}{{Starck} et~al.}{2001}]{2001A&A...368..730S}
{Starck} J.-L.,  {Murtagh} F.,  {Querre} P.,   {Bonnarel} F.,  2001, \mn@doi
  [\aap] {10.1051/0004-6361:20000575}, \href
  {http://cdsads.u-strasbg.fr/abs/2001A%26A...368..730S} {368, 730}

\bibitem[\protect\citeauthoryear{{Starck}, {Pires}  \&
  {R{\'e}fr{\'e}gier}}{{Starck} et~al.}{2006}]{2006A&A...451.1139S}
{Starck} J.-L.,  {Pires} S.,   {R{\'e}fr{\'e}gier} A.,  2006, \mn@doi [\aap]
  {10.1051/0004-6361:20052997}, \href
  {http://adsabs.harvard.edu/abs/2006A%26A...451.1139S} {451, 1139}

\bibitem[\protect\citeauthoryear{{Starck}, {Fadili}  \& {Rassat}}{{Starck}
  et~al.}{2013}]{2013A&A...550A..15S}
{Starck} J.-L.,  {Fadili} M.~J.,   {Rassat} A.,  2013, \mn@doi [\aap]
  {10.1051/0004-6361/201220332}, \href
  {http://adsabs.harvard.edu/abs/2013A%26A...550A..15S} {550, A15}

\bibitem[\protect\citeauthoryear{{Suyu}, {Marshall}, {Auger}, {Hilbert},
  {Blandford}, {Koopmans}, {Fassnacht}  \& {Treu}}{{Suyu}
  et~al.}{2010}]{2010ApJ...711..201S}
{Suyu} S.~H.,  {Marshall} P.~J.,  {Auger} M.~W.,  {Hilbert} S.,  {Blandford}
  R.~D.,  {Koopmans} L.~V.~E.,  {Fassnacht} C.~D.,   {Treu} T.,  2010, \mn@doi
  [\apj] {10.1088/0004-637X/711/1/201}, \href
  {http://adsabs.harvard.edu/abs/2010ApJ...711..201S} {711, 201}

\bibitem[\protect\citeauthoryear{{Suyu} et~al.,}{{Suyu}
  et~al.}{2013}]{2013ApJ...766...70S}
{Suyu} S.~H.,  et~al., 2013, \mn@doi [\apj] {10.1088/0004-637X/766/2/70}, \href
  {http://adsabs.harvard.edu/abs/2013ApJ...766...70S} {766, 70}

\bibitem[\protect\citeauthoryear{{Suyu} et~al.,}{{Suyu}
  et~al.}{2017}]{2017MNRAS.468.2590S}
{Suyu} S.~H.,  et~al., 2017, \mn@doi [\mnras] {10.1093/mnras/stx483}, \href
  {http://adsabs.harvard.edu/abs/2017MNRAS.468.2590S} {468, 2590}

\bibitem[\protect\citeauthoryear{{Takada} \& {Jain}}{{Takada} \&
  {Jain}}{2003}]{2003MNRAS.344..857T}
{Takada} M.,  {Jain} B.,  2003, \mn@doi [\mnras]
  {10.1046/j.1365-8711.2003.06868.x}, \href
  {http://adsabs.harvard.edu/abs/2003MNRAS.344..857T} {344, 857}

\bibitem[\protect\citeauthoryear{{Treu} \& {Marshall}}{{Treu} \&
  {Marshall}}{2016}]{2016A&ARv..24...11T}
{Treu} T.,  {Marshall} P.~J.,  2016, \mn@doi [\aapr]
  {10.1007/s00159-016-0096-8}, \href
  {http://adsabs.harvard.edu/abs/2016A%26ARv..24...11T} {24, 11}

\bibitem[\protect\citeauthoryear{{Utsumi}, {Miyazaki}, {Geller},
  {Dell'Antonio}, {Oguri}, {Kurtz}, {Hamana}  \& {Fabricant}}{{Utsumi}
  et~al.}{2014}]{2014ApJ...786...93U}
{Utsumi} Y.,  {Miyazaki} S.,  {Geller} M.~J.,  {Dell'Antonio} I.~P.,  {Oguri}
  M.,  {Kurtz} M.~J.,  {Hamana} T.,   {Fabricant} D.~G.,  2014, \mn@doi [\apj]
  {10.1088/0004-637X/786/2/93}, \href
  {http://adsabs.harvard.edu/abs/2014ApJ...786...93U} {786, 93}

\bibitem[\protect\citeauthoryear{{Vale}, {Hoekstra}, {van Waerbeke}  \&
  {White}}{{Vale} et~al.}{2004}]{2004ApJ...613L...1V}
{Vale} C.,  {Hoekstra} H.,  {van Waerbeke} L.,   {White} M.,  2004, \mn@doi
  [\apjl] {10.1086/424873}, \href
  {http://adsabs.harvard.edu/abs/2004ApJ...613L...1V} {613, L1}

\bibitem[\protect\citeauthoryear{{Vanderriest}, {Schneider}, {Herpe},
  {Chevreton}, {Moles}  \& {Wlerick}}{{Vanderriest}
  et~al.}{1989}]{1989A&A...215....1V}
{Vanderriest} C.,  {Schneider} J.,  {Herpe} G.,  {Chevreton} M.,  {Moles} M.,
  {Wlerick} G.,  1989, \aap, \href
  {http://adsabs.harvard.edu/abs/1989A%26A...215....1V} {215, 1}

\bibitem[\protect\citeauthoryear{{Vikram} et~al.,}{{Vikram}
  et~al.}{2015}]{2015PhRvD..92b2006V}
{Vikram} V.,  et~al., 2015, \mn@doi [\prd] {10.1103/PhysRevD.92.022006}, \href
  {http://adsabs.harvard.edu/abs/2015PhRvD..92b2006V} {92, 022006}

\bibitem[\protect\citeauthoryear{{Wilson}, {Zabludoff}, {Ammons}, {Momcheva},
  {Williams}  \& {Keeton}}{{Wilson} et~al.}{2016}]{2016ApJ...833..194W}
{Wilson} M.~L.,  {Zabludoff} A.~I.,  {Ammons} S.~M.,  {Momcheva} I.~G.,
  {Williams} K.~A.,   {Keeton} C.~R.,  2016, \mn@doi [\apj]
  {10.3847/1538-4357/833/2/194}, \href
  {http://adsabs.harvard.edu/abs/2016ApJ...833..194W} {833, 194}

\bibitem[\protect\citeauthoryear{{Wilson}, {Zabludoff}, {Keeton}, {Wong},
  {Williams}, {French}  \& {Momcheva}}{{Wilson}
  et~al.}{2017}]{2017arXiv171009900W}
{Wilson} M.~L.,  {Zabludoff} A.~I.,  {Keeton} C.~R.,  {Wong} K.~C.,  {Williams}
  K.~A.,  {French} K.~D.,   {Momcheva} I.~G.,  2017, preprint, \href
  {http://adsabs.harvard.edu/abs/2017arXiv171009900W} {} (\mn@eprint {arXiv}
  {1710.09900})

\bibitem[\protect\citeauthoryear{{Wisotzki}, {Christlieb}, {Bade}, {Beckmann},
  {K{\"o}hler}, {Vanelle}  \& {Reimers}}{{Wisotzki}
  et~al.}{2000}]{2000A&A...358...77W}
{Wisotzki} L.,  {Christlieb} N.,  {Bade} N.,  {Beckmann} V.,  {K{\"o}hler} T.,
  {Vanelle} C.,   {Reimers} D.,  2000, \aap, \href
  {http://adsabs.harvard.edu/abs/2000A%26A...358...77W} {358, 77}

\bibitem[\protect\citeauthoryear{{Wisotzki}, {Schechter}, {Bradt},
  {Heinm{\"u}ller}  \& {Reimers}}{{Wisotzki}
  et~al.}{2002}]{2002A&A...395...17W}
{Wisotzki} L.,  {Schechter} P.~L.,  {Bradt} H.~V.,  {Heinm{\"u}ller} J.,
  {Reimers} D.,  2002, \mn@doi [\aap] {10.1051/0004-6361:20021213}, \href
  {http://adsabs.harvard.edu/abs/2002A%26A...395...17W} {395, 17}

\bibitem[\protect\citeauthoryear{{Wong} et~al.,}{{Wong}
  et~al.}{2017}]{2017MNRAS.465.4895W}
{Wong} K.~C.,  et~al., 2017, \mn@doi [\mnras] {10.1093/mnras/stw3077}, \href
  {http://adsabs.harvard.edu/abs/2017MNRAS.465.4895W} {465, 4895}

\bibitem[\protect\citeauthoryear{{van Waerbeke}}{{van
  Waerbeke}}{2000}]{2000MNRAS.313..524V}
{van Waerbeke} L.,  2000, \mn@doi [\mnras] {10.1046/j.1365-8711.2000.03259.x},
  \href {http://adsabs.harvard.edu/abs/2000MNRAS.313..524V} {313, 524}

\makeatother
\end{thebibliography}
\bibliographystyle{mnras}


\bsp	
\label{lastpage}
\end{document}